\documentclass[prb,amsmath,amsfonts,amssymb,twocolumn,superscriptaddress,showpacs]{revtex4-1}

\usepackage{graphicx,psfrag,color}% Include figure files
\usepackage{dcolumn}% Align table columns on decimal point
\usepackage{bm}% bold math
\usepackage{mathbbol,verbatim}

\begin{document}

\title{A dynamical relation between dual finite temperature classical and zero temperature quantum systems: quantum critical jamming and quantum dynamical heterogeneities}

\author{Zohar Nussinov}
\affiliation{Department of Physics, Washington University, St.
Louis, MO 63160, USA}
\affiliation{Kavli Institute for Theoretical Physics, Santa Barbara, CA 93106, USA}
\author{Patrick Johnson}
\affiliation{Department of Physics, Washington University, St.
Louis, MO 63160, USA}

\author{Matthias J. Graf}
\affiliation{Theoretical Division, Los Alamos National Lab, NM 87545, USA}
\author{Alexander V. Balatsky}
\affiliation{Theoretical Division, Los Alamos National Lab, NM 87545, USA}
\affiliation{NORDITA, Roslagstullsbacken 23, 106 91 Stockholm, Sweden}

\date{\today}

\begin{abstract}
  Many electronic systems exhibit striking features in their dynamical response over a prominent range of experimental parameters. While there are empirical suggestions of particular increasing length scales that accompany such transitions, this identification is not universal. To better understand such behavior in quantum systems, we extend a known mapping (earlier studied in stochastic, or supersymmetric, quantum mechanics) between finite temperature classical Fokker-Planck systems and related quantum systems at zero temperature to include general non-equilibrium dynamics. Unlike Feynman mappings or stochastic quantization methods (or holographic type dualities), the classical systems that we consider and their quantum duals reside in the same number of space-time dimensions. The upshot of our exact result is that a Wick rotation relates (i) dynamics in general finite temperature classical dissipative systems to (ii) zero temperature dynamics in the corresponding dual many-body quantum systems.  Using this correspondence, we illustrate that, even in the absence of imposed disorder, many continuum quantum fluid systems (and possible lattice counterparts) may exhibit a zero-point ``quantum dynamical heterogeneity'' wherein the dynamics, at a given instant, is spatially non-uniform. While the static length scales accompanying this phenomenon do not exhibit a clear divergence in standard correlation functions, the length scale of the dynamical heterogeneities can increase dramatically. We study ``quantum jamming'' and illustrate how a hard core bosonic system may undergo a zero temperature quantum critical metal-to-insulator-type transition with an extremely large effective dynamical exponent $z>4$ consistent with length scales that increase far more slowly than the relaxation time as a putative critical transition is approached. We suggest ways to analyze experimental data. 
\end{abstract}

\maketitle

%%%%%%%%%%%%%%%%%%%%%%%%%%%%%%%%%%%%%%%%%%%%%%%%%%%%%%%%%%%%%%%%%%%%%
%%% Intro

\section{Introduction}

A prominent centerpiece in the understanding of numerous systems
is Landau Fermi liquid (LFL) theory; this theory allows the understanding
of phenomena such as conventional metals and low temperature $^{3}$He liquids. 
LFL theory is centered on the premise that the low energy states of
interacting electron systems may be captured by long-lived fermionic quasi-particles 
with renormalized  parameters (e.g., effective masses that differ from those of the bare
electron).  The last three decades have seen the discovery of materials in which 
electronic behavior deviates from simple LFL type behavior.  These ``singular'' or ``non-Fermi liquids'' 
(NFL) include the pseudo-gap region of the high-temperature cuprate superconductors and ``heavy fermions'' 
(in which, as befits their name, the effective electron mass becomes very large). 
While there are clear indications of changes in the dynamics in these systems--including 
putative quantum critical points \cite{subir_book,physics_reports} -- there is, in most cases, no clear experimentally measured 
length scale that exhibits a clear divergence. A quantum critical point is associated with a continuous phase transition at (absolute) 
zero temperature. 
Typically, this may occur in a system whose transition temperature is driven to zero by doping 
or the application of magnetic fields or pressure.
Within a quantum critical regime, response functions follow universal power law scaling 
in both space and time. Specifically, at a quantum critical point, the effective infrared (IR) fixed point  theory 
exhibits scaling invariance in space-time: $t\to\lambda t, ~\vec{x}\to\lambda^{1/z} \vec{x}$ with a dynamical exponent $z$ that
can, depending on the theory at hand, assume various canonical values. 
Unlike classical critical points whose associated critical fluctuations are confined 
to a narrow region near the phase transition, quantum critical fluctuations appear over 
a wide range of temperatures above the quantum critical point.
These fluctuations may generally lead to a radical departure of the system's electronic 
properties from standard LFL type behavior.
These features are anticipated to be common across many strongly correlated electronic 
systems and may be associated, in some electronic systems, with a change of Fermi surface 
topology \cite{top}. The genesis of NFL behavior in myriad systems has attracted much attention.
Various theoretical proposals for quantum critical points in NFL, include, amongst many others, those
that raise the specter of new special topological excitations \cite{aji}.
In the current work, we wish to suggest that some aspects of effectively local behaviors
exhibited by many strongly correlated electronic systems might rather be 
understood as direct quantum renditions of known classical behaviors. 

Many NFL systems exhibit numerous phases (including, quite notably, superconductivity).
Indeed, competing orders and proliferation of multiple low energy states can lead to spatially non-uniform glassy characteristics \cite{nvb}
and associated first order quantum transitions \cite{She}. 
The length scales characterizing these electronic systems undergo a much milder change than the
corresponding changes in the dynamics.
All of this suggests an effective infrared (IR) fixed point quantum field theory is invariant under 
scaling in time but not in space- i.e., the effective 
dynamical exponent $z \to \infty$.
Response functions such as those of the marginal Fermi liquid form describing cuprates near optimal doping
show a marked frequency dependence but essentially no spatial momentum dependence.
In this work, we will suggest how many quantum systems might exhibit very large effective dynamical exponents.

It is natural to first look elsewhere in physics where similar phenomena appear. One arena immediately comes to mind.
In classical structural glasses there is a dramatic change in the dynamics 
as a liquid is rapidly cooled (supercooled) below its melting temperature it falls out of equilibrium at low temperatures and becomes quenched into a glass without the appearance of easily discernible large 
changes in measurable standard static length scales. 
While the ergodicity breaking that accompanies a glass transition cannot occur in a finite 
size system, it essentially mandates the appearance of a diverging static length scale \cite{montanari},
but such divergent length scales generally do not simply manifest themselves in bare standard correlation functions.
General correlation functions which may monitor subtle changes include the ``point to set'' 
\cite{BiroliPointToSet} correlations and others. 
Practically, in most instances \cite{tanakaorder} %(***orientational order of Tanaka**) 
no clear signatures of divergent length scales are easily seen in standard static two-point correlation functions. 

A far more transparent growth in length scales is seen from four-point correlation functions that
quantify the change in correlations as the system evolves in time. 
These correlation functions afford a glimpse into the length and time scaling
which describe dynamical heterogeneities that characterize the spatially
non-uniform rate of change or dynamics in the system.
The length scale associated with these heterogeneities was seen
to grow as the characteristic relaxation times increased. 

We may use similar correlation 
functions to characterize strongly correlated electronic systems in which there
are 
strongly discernible changes 
in the dynamics but no obvious experimentally 
accessible tools that point to accompanying divergent length scales in a general 
way \cite{LawlerNematicOrder}. 
To our knowledge, to date, dynamical heterogeneities (or general static measures 
such as those of the point-to-set method) have not been systematically probed in electronic systems
nor has their existence been established as a matter of principle in quantum system.
Initial ideas concerning non-uniform doping-driven heterogeneities were discussed in \cite{papa}.
In this work, we flesh out the blueprint for a proof outlined, by one of us, in \cite{physics_ZN} and
we will provide concrete ``matter of principle''
theoretical testimony to the emergence of quantum dynamical heterogeneities in {\it clean systems} and related properties in quantum many body systems. 
Some time after \cite{physics_ZN}, the authors of \cite{olmos} have further confirmed the existence of quantum dynamical heterogeneities
in certain dissipative spin systems. In what follows, we focus on systems such as structural glasses
or jammed systems with no external disorder. 

As is well appreciated, such translationally invariant systems that are free of external disorder may, on their own, display non-uniform spatial patterns concomitant with interesting dynamical properties \cite{htbib}. 
For times far shorter than the equilibration times, the two-point auto-correlation function,
\begin{equation}
  C(\vec{x},t) = \langle  \delta \phi(\vec{x},t) \delta \phi(\vec{x},0) \rangle,
  \label{ct}
 \end{equation}
 with the brackets denoting an average over a translationally invariant equilibrium average,
 records dynamic fluctuations at all points $\vec{x}$ in $d$ dimensional space. In Eq. (\ref{ct}), the (deviations of the) fields $\delta \phi(\vec{x},t)$ may correspond to, e.g., density or any other pertinent spatio-temporal quantity characterizing the system. A notable variant in classical liquids is that for the mobility field
 where, in all correlators, $\delta \phi (\vec{x},t)$ is replaced by
 \begin{eqnarray}
 \label{mobility}
\mu (\vec{x}, t) =  \sum_{i=1}^{N}  w_{i}(t) \delta[\vec{x}- \vec{x}_{i}(0)],
 \end{eqnarray}
with an individual particle ($i$) mobility $w_{i}$ monitoring particle motion during a time interval of size $t$, e.g.,  $w_{i}(t) = \exp(|\vec{r}_{i}(t) - \vec{r}_{i}(0)|^{2}/d^{2})$ with $d$ the particle size.
The two-point correlator of Eq. (\ref{ct}) constitutes an analogue of the Edwards-Anderson \cite{EA} order parameter that appears in spin glasses.
In uniform systems, the correlator of Eq. (\ref{ct}) is spatially ($\vec{x}$) independent. 
The spatial correlation amongst pair products of time separated products of fields (such as those in Eq. (\ref{ct}))
at different spatial sites is a four-point correlation function  \cite{4p}
that attempts to measure cooperation
\begin{eqnarray}
  G_{4}(\vec{x}-\vec{y},t) = \langle \delta \phi(\vec{x},t) \delta \phi(\vec{x},0) 
                         \delta \phi(\vec{y},t) \delta \phi(\vec{y},0) \rangle \nonumber
                         \\ - C(\vec{x},t) C(\vec{y},t) .
                         \label{g4xy}
\end{eqnarray}
This four-point correlation function relates the dynamics at two different spatial points $\vec{x}$ and $\vec{y}$. Generally, in the absence of quenched disorder, 
due to translational invariance, this correlation function depends only on $(\vec{x}-\vec{y})$ and not on $\vec{x}$ and $\vec{y}$ separately.
Empirically, the integral of this correlation function,
\begin{eqnarray}
\chi_{4}(t) =  \int d^{d} x ~G_{4}(\vec{x},t) = \langle \tilde{C}^{2}(t) \rangle   -  \langle \tilde{C}(t) \rangle^{2},
                         \label{g4xy+0}
\end{eqnarray}
where $\tilde{C}(t) = \int d^{d} x ~  \delta\phi(\vec{x},t) \delta \phi(\vec{x},0)$ 
(or, correspondingly, $\tilde{C}(t) = \int d^{d}x ~ \mu(\vec{x}, t) \mu(\vec{x},0)$ for spatio-temporal correlations of the mobility of Eq. (\ref{mobility}))
is often employed to enable a quantification of dynamical heterogeneities
when the atomic coordinates may be resolved \cite{htbib}.

\section{Outline}
This article illustrates that the above mentioned zero-point dynamical heterogeneities can indeed rear their head in quantum many body systems. It more rigorously shows that zero-temperature quantum many body systems
can, apart from quantum critical phenomena \cite{subir_book,physics_reports}, exhibit exact analogs of finite temperature classical behavior including that in 
glass forming systems. Towards this end, we establish two sets of results:

(1)  Simple mathematical relations between the dynamical correlation functions in (different) classical and quantum systems when these systems are linked to one another by a duality that we will describe. 

(2) Physical consequences of these mathematical relations when these are applied to classical glass forming systems (whence they lead to
predictions for the zero temperature time dependent correlation functions in dual quantum systems).

 Accordingly, our work is split into, roughly, two equal parts describing physical background and derivations that realize both of these endeavours. 
 The outline of this paper is as follows. In section \ref{rev-sec}, we establish our main result  
 given by Eq. (\ref{qcqc}), which links time dependent correlation functions in 
 finite temperature dissipative classical systems with their dual zero temperature quantum systems. 
 After stating our result,
 much time is spent deriving it from first principles. Towards this end, we first review the rudiments of stochastic quantum
 mechanics and then derive in detail the key result in Eq. (\ref{qcqc})
 and similar relations like it for higher order correlations.
 In section \ref{models}, we explicitly list several (of the many) dissipative classical systems that exhibit dynamical heterogeneities 
 and other facets of glassy dynamics and their quantum duals. 
 
 With these relations in tow, we then proceed to discuss physical predictions for quantum many body systems. 
 In section \ref{glassy}, we establish the existence of quantum dynamical heterogeneities, scaling
 of relaxation times, and quantum critical jamming. In the quantum dual models
 that exhibit these phenomena, the relaxation times increase much
 more rapidly than correlation length scales.
Further building on these results, in sections \ref{ls}, \ref{esp} we introduce hard-core Bose systems as well
as electronic systems with pairing interactions that may display glassy dynamics. 
We outline in section \ref{exp_implications} data analysis that may validate
the presence of quantum dynamical heterogeneities in experimental systems. 
We conclude (section \ref{conc}) with a synopsis of our central results.
Several technical details have been relegated to the appendices
(including, perhaps most notably (appendix \ref{dets})
the proper analytic continuation for 
stretched exponential dynamical correlations in 
dissipative classical systems to obtain correlations
in the corresponding quantum glass systems).

\section{A dynamical relation between viscous classical and quantum many body systems}
\label{rev-sec}
In order to illustrate how, as a matter of principle, the physics of such classical 
dissipative systems can appear in clean quantum systems at zero temperature, we employ and extend
a mapping \cite{Parisi,Zinn,fenyes,Nelson,Guerra1,Guerra2,Biroli,castel,fei} between classical dissipative systems and quantum many body systems
to include general dynamical (including non-equilibrium) systems. Aspects of this mapping are intimately linked to Madelung hydrodynamics \cite{madelung}
which links the real and imaginary parts of the Schr\"odinger equation to classical hydrodynamics. 
Related work concerning dynamics in Rokshar-Kivelson \cite{RK*} systems
also appears in \cite{RK}. Below, we first briefly review this mapping. We will then derive a hitherto unknown result linking the dynamics in
these classical and quantum systems. %have been relegated to Appendix \ref{TwoOperators}.

The crux of the``stochastic quantum mechanics''  mapping between dissipative classical systems and many body bosonic theories  \cite{Parisi,Zinn,fenyes,Nelson,Guerra1,Guerra2,Biroli,castel,fei},
is the realization that the equation of motion for 
a dissipative (or ``Aristotelian'') classical system is a first order differential equation in time just as
the Schr\"{o}dinger equation is. Using this equivalence, systems obeying the Langevin equation,
\begin{equation}
  \gamma_{i} \frac{d\vec{x}_{i}}{dt} = - \vec{\nabla}_{i} 
                                 V_{\sf N}(\vec{x}_{1}, ..., \vec{x}_{N}) + \vec{{\eta}}_{i}(t),
  \label{Langevin}
\end{equation}
with $i$ the particle index, $\gamma_{i}$ the coefficients of friction, 
$\eta_{i}^{\alpha}(t)$ the Gaussian noise with 
\begin{eqnarray}
\label{noiseeq}
\langle \eta_{i}^\alpha(t) \eta_{j}^{\beta}(t') \rangle = 2T_{cl} \gamma_{i} \delta_{ij} \delta_{\alpha \beta} \delta(t-t'),
\end{eqnarray}
 where
$T_{cl}$ is the effective temperature of the classical system, and, $\alpha, \beta = 1, 2,\ldots , d$  
can be exactly mapped \cite{Parisi,Zinn,Biroli} onto a quantum many body system of bosons with effective mass $m_{i} = \gamma_{i}/(2T_{cl})$
at zero temperature which is governed by the Hamiltonian 
\begin{eqnarray}
\label{HQM}
  H & = & \sum_{i} \frac{1}{\gamma_{i}} \left[-T_{cl} %\frac{\partial^{2}}{\partial \vec{x}
  \nabla_{i}^{2}
          - \frac{1}{2} \nabla^{2}_{i} V_{\sf N} + \frac{1}{4T_{cl}} (\nabla_{i} V_{\sf N})^{2}\right] \nonumber\\ 
    & \equiv & \sum_{i} \frac{p_{i}^{2}}{2m_{i}} + {\cal{V}}_{\sf Quantum}(\{\vec{x}\}).
\end{eqnarray}
We will term the quantum system of Eq. (\ref{HQM}) ``the dual quantum system'' associated with the 
classical system of Eqs. (\ref{Langevin}, \ref{noiseeq}).
The many body quantum potential ${\cal{V}}_{\sf Quantum}(\{\vec{x}\})$ is constructed from 
the gradients of the classical potential energy $V_{N}$ as in Eq. (\ref{HQM}).  
Under this mapping, \cite{Parisi,Zinn,fenyes,Nelson,Guerra1,Guerra2} a dissipative classical system with a potential energy $V_{N}$ that captures {\it repulsive hard core spheres} maps onto a quantum system
at zero temperature with (as is apparent in the many body potential energy ${\cal{V}}_{\sf Quantum}$) similar {\it dominant {\it hard-core} interactions} (augmented by soft sticky interactions) \cite{Biroli}.
Although we will focus on the mapping from
classical systems to corresponding quantum ones, it is also possible to go in the opposite direction and map 
quantum mechanical systems with known non-degenerate ground states onto classical dissipative systems (see appendix \ref{aq2c}).
(For completeness, we note that different ``stochastic quantization'' \cite{SQ} mappings relate stochastic systems to quantum field theories by the introduction of an additional fictitious time coordinate. An additional time
like coordinate also appears in well known textbook type Feynman mapping as well as far more recent holographic dualities \cite{holo}. By contrast,
in the ``stochastic quantum mechanics'' mappings between classical to quantum systems that we review and expand on here to generally include dynamics, the number of space-time dimensions is identical.)

Earlier work \cite{Parisi,Zinn,Biroli,fenyes,Nelson,Guerra1,Guerra2} advanced the rudiments of this mapping and further suggested dynamical aspects that might follow from it.
  In the current work, we explicitly derive and prove a general and rather powerful relation between arbitrary correlation functions in dissipative classical systems with 
  time varying potentials (necessary for our discussion of quenching) and relate these to corresponding correlation functions
  in zero-temperature quantum systems. In subsection \ref{TwoOperators}, we summarize, following \cite{Biroli,Parisi,fenyes,Nelson,Guerra1,Guerra2}, more detailed aspects of this mapping.  
We now proceed to set the stage for our result and its consequences. 
In what follows, we consider a general classical two-point correlation function of the form
\begin{eqnarray}
G_{\sf classical}(t) = \langle {\cal{O}} (t) {\cal{O}}(0) \rangle,
\label{gclq}
\end{eqnarray}
where ${\cal{O}}(t)$ is {\em any} quantity, at times $t \ge 0$. Our mapping covers general non-equilibrium time dependent Hamiltonians
in which only the initial (or only the final) classical system is in thermal equilibrium at an initial (or final) temperature.  
In such a case, the corresponding dual quantum problem evolves unitarily with a time dependent Hamiltonian. 
Specifically, as we will elaborate on in subsection \ref{TwoOperators}, we find a very simple and general result 
for {\it any} quantity ${\cal{O}}(t)$
\begin{eqnarray}
\label{qcqc}
\boxed{G_{\sf Quantum} (t) = G_{\sf classical}(it)}.
\end{eqnarray}
Thus,
a ``complexification'' of the time coordinate (or {\it Wick-type rotation}) relates general classical dynamical correlation functions
of the form of Eq. (\ref{gclq}) to their corresponding quantum counterparts. The quantum correlation function $ G_{\sf Quantum}(\vec{q},t) $ is evaluated with 
the quantum many body potential ${\cal{V}}_{\sf Quantum}(\{\vec{x}\})$ of Eq. (\ref{Potential}), while the corresponding classical correlation function is computed for a system
with a potential $V_{\sf classical}(\{\vec{x}\})$.
%wherein for $t \ge 0$,
Similarly, for the quantum linear response functions $R_{\sf Quantum}(t)$ [see Eqs. (\ref{LinearInFofT}, \ref{rquantum}) for standard linear response expressions], we have 
that
\begin{eqnarray}
R_{\sf Quantum}(t) = i  \Theta(t) (G_{\sf classical}(it) - G^{*}_{\sf classical}(it)).
\label{ggqc}
\end{eqnarray}

Eqs. (\ref{qcqc},\ref{ggqc}) provide an entry in the mapping between the finite temperature classical system of Eq. (\ref{Langevin}) governed by a 
potential $V_{\sf N}$ and its corresponding quantum zero-temperature dual 
with a quantum many body potential energy $ {\cal{V}}_{\sf Quantum}(\{\vec{x}\})$
in the Hamiltonian of Eq. (\ref{HQM}).

Eqs. (\ref{qcqc},\ref{ggqc}), along with their consequences are key results of this work. In subsection \ref{TwoOperators}, we provide a detailed derivation of Eqs. (\ref{qcqc},\ref{ggqc}). 
Typically, in glassy systems, the correlation function of Eq. (\ref{gclq}) is a superposition of many decaying modes.
This distribution of modes will manifest itself as a distribution of oscillatory modes in the corresponding dual quantum problem.
In many cases, this will lead to zero temperature quantum dynamics of the dual system that, with additional oscillations, will emulate the finite classical dynamics.
For instance, as it precisely occurs in viscous systems with overdamped dynamics (for which Eq. (\ref{Langevin}) applies), if for times $t>0$,
\begin{eqnarray}
\label{stretch}
G_{\sf classical}(t) =A \exp[- (t/ \tau)^{c}]
\end{eqnarray} 
then, correspondingly for all of these positive times $t$ (see Appendix \ref{dets}),
\begin{eqnarray}
\label{qqq}
R_{\sf Quantum}(t) = 2A e^{-(\frac{t}{\tau})^{c} \cos \frac{\pi c}{2}}
 \sin\Big[\left(\frac{t}{\tau}\right)^{c}\sin 
\frac{\pi c}{2}
\Big].
\end{eqnarray}
With the aid of the general relation of Eqs. (\ref{qcqc},\ref{ggqc}), the quantum correlation function that corresponds to a general stretched exponential correlation function in the classical arena 
can be computed analytically. The linear response function is, indeed, given by Eq.(\ref{qqq}). A trivial yet important particular corollary of Eq. (\ref{qcqc}) is that static correlations (i.e., those for $t=0$)
are identical in the finite temperature classical and their corresponding quantum dual systems. 

The remainder of this section is organized as follows. In subsection \ref{FPreview}, we review the basic known essentials of the mapping between finite temperature classical dissipative 
systems and their quantum duals. This is then followed, in subsection \ref{TwoOperators}, by the derivation of our result of Eq. (\ref{qcqc}).

\subsection{Lightning review of known relations in stochastic quantum mechanics}
\label{FPreview}

This subsection reviews earlier work which is necessary for our derivations in the subsections that follow. We will now briefly highlight known salient features of the mapping \cite{Parisi,Zinn,Biroli,fenyes,Nelson,Guerra1,Guerra2} 
between classical stochastic systems and their quantum duals (aka ``stochastic quantum mechanics'').
The subsections (and sections) that follow will build on the classical to quantum mapping that we now discuss here.
There are two prominent independent approaches that establish this duality: (i) a general method that examines the Fokker-Planck equations
and (ii) an algebraic method highlighting a harmonic oscillator (with a simple raising/lowering operator) type structure
akin to that more generally found in supersymmetric theories. Both approaches directly lead to the effective quantum Hamiltonian of
Eq. (\ref{HQM}). Although this Hamiltonian is more readily seen in the algebraic formulation and leads to immediate and clear intuition
(which is why we briefly review it), the broader approach is arguably that relying
on the direct Fokker-Planck evolution of dissipative classical systems. It is this Fokker-Planck approach that we will use in our derivations
in the subsections that follow.

\subsubsection{Fokker Planck systems and their quantum duals}
\label{FPqs}

We first set the preliminaries for the Fokker Planck approach following \cite{Zinn}. Given an initial vector $x_{0}$ of the coordinates of all particles at time $t=t_{0}$,
the time dependent probability distribution ${\cal{P}}(\vec{x},t;\vec{x}_{0}, t_{0})$ for
the corresponding position vectors $\{\vec{x}(t)\}$ at time $t$
is given by
\begin{eqnarray}
\label{deltaP}
{\cal{P}}(\{\vec{x}\},t;\{\vec{x}_{0}\},t_{0}) = \langle \prod_{i=1}^{N} \delta[\{\vec{x}_{i}(t)-\vec{x}_{i}\}]
\rangle_{\{\vec{\eta}\},\{\vec{x}_{0}\}},
\end{eqnarray}
where $\langle - \rangle_{\eta,\{\vec{x}_{0}\}}$ denotes the average over the random noise $\eta$
(which we will take to be the Gaussian white noise of Eq. (\ref{noiseeq})) 
given that initially, at time $t=t_{0}$, the particle coordinates were $\{\vec{x}_{0}\}$.
The average of a general function ${\cal{O}}(\{\vec{x}(t)\})$ is
then
\begin{eqnarray}
\int d^{dN}x {\cal{P}}(\{\vec{x}\},t;\{\vec{x}_{0}\},t_{0}) {\cal{O}}(\{\vec{x}\}) \nonumber
\\  = \langle {\cal{O}}(\{\vec{x}(t)\})
\rangle_{\{\vec{\eta}\},\{\vec{x}_{0}\}}.
\end{eqnarray}

It is convenient to write ${\cal{P}}(\{\vec{x}\},t; \{\vec{x}_{0}\},t_{0})$ in a Dirac notation as
\begin{eqnarray}
{\cal{P}}(\{ \vec{x} \},t; \{ \vec{x}_{0}\},t_{0}) = \langle \{ \vec{x} \}| \hat{P}(t,t_{0}) | \{ \vec{x}_{0}\} \rangle.
\end{eqnarray}
Time translation invariance and the Markov property of these probabilities,
\begin{eqnarray}
\int d^{dN}x'  \langle \{ \vec{x} \}| \hat{P}(t,t') | \{ \vec{x}' \} \rangle \langle \{ \vec{x}' \}| \hat{P}(t',t_{0}) |
\{ \vec{x}_{0} \} \rangle \nonumber
\\ =  \int d^{dN}x' {\cal{P}}(\{\vec{x}\},t;\{ \vec{x}'\},t') {\cal{P}}(\{ \vec{x}' \},t';\{ \vec{x}_{0}\},t_{0}) \nonumber
\\ =
{\cal{P}}(\{ \vec{x} \},t; \{ \vec{x}_{0} \},t_{0}),
\end{eqnarray}
imply that
\begin{eqnarray}
\label{h't}
\hat{P}(t,t_{0}) ={\cal{T}} e^{-\int_{t_0}^{t}H_{FP}(t')dt'},
\end{eqnarray}
with a ``Fokker-Planck'' Hamiltonian $H_{PF}$  and where ${\cal{T}}$ is the time ordering operator.
We now return to our particular classical to quantum mapping.
The summary below closely follows this mapping as presented by Biroli et al. \cite{Biroli}. 
In what follows, we set
\begin{eqnarray}
\label{pschr}
P=\hat{P}(t,t_{0}) | \{ \vec{x}_{0}\} \rangle.
\end{eqnarray}
For the classical dissipative system of Eq. (\ref{Langevin}), the probability distribution $P(\{\vec{x}\})$ evolves according to the Fokker-Planck equation
\begin{equation}
\label{FPEQ}
  \frac{\partial P}{\partial t} = -H_{FP} P,
\end{equation}
where the Fokker Planck operator is
\begin{equation}
  H_{FP} = - \sum_{i} \frac{1}{\gamma_{i}} \vec{\nabla}_{i} \cdot
           \left[ \vec{\nabla}_{i} V_{\sf N} + T_{cl} \vec{\nabla}_{i} \right],
  \label{FPOperator}
\end{equation}
with $T_{cl}$ being the temperature of the classical system [setting the noise strength in Eq. (\ref{noiseeq})].
Eq. (\ref{FPOperator}) follows from a direct differentiation of Eq. (\ref{deltaP}) while invoking Eq. (\ref{Langevin}) for the derivatives of the coordinates $\{\vec{x}_{i}(t)\}$ in the argument of the delta functions
and performing short time averages.  (Thus, this equation and our results pertain to systems in which the dynamics is sufficiently slow such that short time averages over the noise $\eta$
at fixed temperature are sensible.) A detailed derivation of the Fokker-Planck equation for this and more general Langevin processes appears in many excellent textbooks, e.g.,  \cite{FPB}. 
The operator $H_{FP}$ is non-Hermitian. Each eigenvalue is generally associated with differing left and right eigenvectors. 
The Fokker-Planck equation can be mapped into a Hermitian Hamiltonian by \cite{feld}
\begin{equation}
\label{QMAC}
  H = e^{V_{\sf N}/(2T_{cl})} H_{FP} e^{-V_{\sf N}/(2T_{cl})} ,
\end{equation}
if the second derivatives are exchangeable, $\vec{\nabla}_{i} \vec{\nabla}_{j }V_{\sf N} = \vec{\nabla}_{j} \vec{\nabla}_{i} V_{\sf N}$.\cite{Risken}
A direct substitution leads to the quantum many body Hamiltonian of Eq. (\ref{HQM}). 
Note that, thus far, we have allowed ${V_{\sf N}}$ to be completely general. This potential energy
may include one body interactions (i.e., coupling to an external source),
pair interactions between particles, and three- and higher-order particle
interactions. A key point that we will further invoke later is that the transformation of Eq. (\ref{QMAC}) 
leading to a Hermitian quantum Hamiltonian
can be trivially performed at any given time slice when $V_{\sf N}$ and $T_{cl}$ 
are, generally, time dependent. It is worth highlighting that in non-equilibrium time dependent classical systems, the temperature $T_{cl}$ is set by the time dependent noise amplitude (following Eq. (\ref{noiseeq})).
 For general $V_{\sf N}(\{\vec{x}\})$, the Fokker-Planck operator of Eq. (\ref{FPOperator}) 
has non-negative eigenvalues \cite{Parisi,Zinn}. For any time independent $H_{FP}$, the zero eigenvalue state---i.e., the ground state---which 
according to Eq. (\ref{FPEQ}) corresponds to a stationary (time independent) probability distribution $P$.
This
is the equilibrium Boltzmann distribution
\begin{eqnarray}
\label{PZ}
P^{equil} (\{ \vec{x}\},t) = \frac{1}{Z_{N}} e^{-\beta V_{\sf N}(\{ \vec{x}\})},
\end{eqnarray}
with $Z_{N}$ the partition function associated with $V_{\sf N}(\{\vec{x}\})$
and $\beta = 1/T_{cl}$. This is readily rationalized by the following argument.
For a finite size system,
the linear eigenvalue equation
\begin{eqnarray}
(H_{FP})_{bc} P_{c} = - \varepsilon_{c} P_{c},
\label{abcp}
\end{eqnarray}
with the matrix row/column indices $b$ and $c$  denoting classical configurations, has a (finite size) matrix $H_{FP}$ with positive off-diagonal elements and negative diagonal entries.
Specifically, in Eqs. (\ref{FPEQ},\ref{abcp}), the transition matrix $H_{FP}$ has entries that relate the probabilities of going from state  
$b$ to state $c$ in a given (infinitesimal) time interval. If these states are different ($b \neq c$) then clearly $(H_{FP})_{bc} >0$. 
The diagonal elements $(H_{FP})_{bb}$ provide the probabilities of  ``leaking out''
of state $b$ and going to all other states $c \neq b$. From all of this it follows
that 
\begin{eqnarray}
\label{FPbb}
(H_{FP})_{bb} = - \sum_{b' \neq b} (H_{FP})_{bb'} <0.
\end{eqnarray}
Detailed balance, i.e., the fact that the probability of going from $b$ to $c$ is the same as that of going from $c$ to $b$, asserts that
\begin{eqnarray}
(H_{FP})_{bc} e^{-\beta E_{b}} = (H_{FP})_{cb} e^{-\beta E_{c}}. 
\label{FPbc}
\end{eqnarray}
Eqs. (\ref{abcp}, \ref{FPbc}) illustrate that the Hamiltonian of Eq. (\ref{QMAC}) is Hermitian. 
In this classical system of Eq. (\ref{Langevin}), the energies of the classical states $E_{c}$ are simply given by $V_{\sf N}(\{\vec{x}\})$ evaluated for the classical configurations $c$. 
With the aid of Eqs. (\ref{FPbb}, \ref{FPbc}), it is easy to see that the column vector $P^{equil}_{c} =  Z_{N}^{-1}  \exp(-\beta E_{c})$ (i.e., the distribution of Eq. (\ref{PZ})) is a null eigenvector of Eq. (\ref{abcp}). 
This probability eigenvector corresponds, of course, to the equilibrium Boltzmann distribution. The factor of $Z_{N}^{-1}$ is inserted to ensure normalization of the classical
probabilities (for any eigenstate): $\sum_{c} P_{c} =1$. Now, we can add a constant to the finite dimensional matrix 
\begin{eqnarray}
 H_{FP} \to  H_{FP} - const.  \equiv   H_{FP}^{'},
\label{constant_shift}
\end{eqnarray} to
generate a matrix $(- H_{FP}^{'})$ that has all of its elements positive $(-H_{FP}^{'})_{bc}>0$. Specifically, to this end, in Eq. (\ref{constant_shift} we can choose $const.$ to be any constant larger than the sign inverted smallest
off-diagonal element of $(-H_{FP})$, i.e., $const. > - \min_{b \neq c} \{H_{FP} \}_{bc}$. For such a positive matrix, we can apply the Perron-Frobenius theorem which states that the largest eigenvector of 
$(-H_{FP}^{'})$ is non-degenerate and that eigenvector is the only eigenvector that has all of its elements positive with all other orthogonal eigenvectors having at least one negative element. 
Clearly, all of the eigenvectors of $H_{FP}$ and $H_{FP}^{'}$ are identical
with the corresponding eigenvectors of both operators merely shifted uniformly by a constant. With all of the above in tow, we see that $P^{equil}$ corresponds to the largest eigenvector of 
$(- H_{FP}^{'})$ and is thus also the largest eigenvector of $(- H_{FP})$. For a time independent $H_{FP}$, as $P^{equil}$ was the null eigenvector of $H_{FP}$, it follows that all other eigenvalues of Eq. (\ref{abcp}), $\varepsilon >0$, 
are positive and, according to Eq. (\ref{FPEQ}) and explicit earlier discussions evolve with time as 
$\lim_{t \to \infty} \exp(- \varepsilon t) = 0$.
Thus, physically (as it to be expected)
at long times the system attains its equilibrium configuration of $P^{equil}$. In the corresponding zero-temperature quantum problem, the dominant classical equilibrium state
with a lowest energy. We will thus label it in Appendix \ref{TwoOperators} by $|G \rangle$.  The transformation of Eq. (\ref{QMAC}) relates the operators in the classical Fokker-Planck 
and zero temperature quantum problem to one another. The transformation for the right eigenvectors $P$ of $H_{FP}$, which we explicitly denote below as $| - \rangle_{FP}$, 
to the eigenvectors of the quantum Hamiltonian $H$
is trivially 
\begin{eqnarray}
\label{ssttate}
\label{rightright}
| - \rangle_{FP} \to \exp(-V_{\sf N}/(2T_{cl}))) | - \rangle_{FP} \nonumber
\\ = |- \rangle_{Quantum}. 
\end{eqnarray}
Similarly, the left eigenvectors ($\langle -|_{FP}$) of $H_{FP}$ are to be multiplied by $\exp(V_{\sf N}/(2T_{cl})))$ in order to pass to the left eigenstates of the quantum problem.
(In the quantum problem defined by $H$, the left and right eigenstates are trivially related to each other by Hermitian conjugation.) 
Applying Eq. (\ref{rightright}) to the null right eigenstate of the Fokker Planck Hamiltonian of Eq. (\ref{PZ}), we see that the quantum eigenstate
of $H$ corresponding to this classical equilibrium state is given  by
\begin{eqnarray}
\label{psie}
\Psi_{0} (\{ \vec{x} \}) = \frac{1}{\sqrt{Z_{N}}}   \exp(- \frac{1}{2T_{cl}}   V_{\sf N}(\{\vec{x}\})).
\end{eqnarray}
The prefactor in Eq. (\ref{psie}) is set by the normalization of this quantum state. 
When comparing Eqs. (\ref{PZ},\ref{psie}) to one another, we see that this wavefunction is related to the classical equilibrium probability eigenvector by the appealing relation $\Psi_{0} (\{ \vec{x} \})
= \sqrt{P^{equil}(\{\vec{x}\})}$ . When $ V_{\sf N}(\{\vec{x}\}))$ is symmetric under the interchange of particle coordinates, the resulting wavefunction {\it may describe bosons}. 
 For two body interactions $V_{ij}$ (Eq. \ref{vnc}) that are symmetric under the permutations of $i$ with $j$, the wavefunction  
of Eq. (\ref{psie}) is symmetric under any permutations of the particles. Thus, the ground state wavefunction 
of Eq. (\ref{psie}) is a Jastrow type wavefunction describing a bosonic system. Of course, generally, $ V_{\sf N}(\{\vec{x}\}))$ can include not only
two body terms but also single body contributions (local chemical potentials or fields) as well as three- and higher-body interactions. 
Although obvious, it is worth noting that if $ V_{\sf N}(\{\vec{x}\}))$ (and thus the quantum Hamiltonian $H$) is invariant under
any pairwise permutation $P_{ij}$, i.e., if $[H,P_{ij}]=0$ then the symmetry of the initial wavefunction $\Psi_{0}$ (corresponding to
the classical Boltzmann distribution for a system initially at an equilibrium at temperature $T_{cl}$) does not change 
as the system evolves with time (including general arbitrary $H$ corresponding to classical variations in temperature and other 
parameters).

We next briefly discuss a generalization of Eq. (\ref{QMAC}).
It is, of course, possible to write down a general similarity transformation,
\begin{eqnarray}
\label{QMAC+}
  H' = \tilde{S}^{-1} H_{FP} \tilde{S},
\end{eqnarray}
with a time dependent operator $\tilde{S}$.
Under Eq. (\ref{QMAC+}), the Fokker-Planck equation of Eq. (\ref{FPEQ}) reads
\begin{eqnarray}
\partial_{t} \Psi  = -H' \Psi,
\end{eqnarray}  
where $\Psi= \tilde{S}^{-1} P$ 
with $P$ given by Eq. (\ref{pschr}).   

\subsubsection{An algebraic approach relating the ground state of a dual quantum system to the Boltzmann distribution of a finite temperature classical system}

As is well known, there exists a beautiful link between stochastic classical statistical mechanics and supersymmetric
quantum systems, e.g., \cite{SUSY}. This connection is especially immediate for the ground state
wavefunctions which are of zero energy (as indeed that of Eq. (\ref{psie})). This might lead to the impression that the results that we will derive
using the correspondence between classical dissipative systems and quantum duals are
rather limited and special. Informally, this suggested by some to lead to un-normalizable wavefunctions
if non-constant equilibrium classical states are considered. As we will explicitly elaborate in subsection \ref{TwoOperators}, 
the time evolution operator ${\cal{U}}(t)$ in the corresponding dual quantum problem is unitary and thus
if an initial state is normalized (such as that of Eq. (\ref{psie}) corresponding to an initial classical equilibrium state)
then the quantum state will remain normalized at all positive times [and vice versa for a unitary evolution towards a final
state of the form of Eq. (\ref{psie})]. Although the ground state of the quantum problem may be dismissed as trivial
and special, the relations concerning the time evolution to states that are not of the form of Eq. (\ref{psie})
are not as immediate. These relations concerning the dynamics form the core of this work.
For pedagogical purposes, we very briefly review here some central notions concerning the mapping
of the Fokker Planck process to supersymmetric quantum mechanics as they, in particular, pertain to the equilibrium
problem. The explicit use of supersymmetry will not be invoked in the below and the discussion
will be made as simple as possible. The Hamiltonian of Eq. (\ref{HQM}) can, for fixed $\gamma_{i} = \gamma$ in a simple (single-particle) one-dimensional rendition
which we adopt for ease of notation, be written as
\begin{eqnarray}
\label{ssy}
H =  \frac{T_{cl}}{\gamma} A^{\dagger} A
\end{eqnarray}
where, 
\begin{eqnarray}
\label{saa}
A^{\dagger} = -  \frac{\partial}{\partial x} + \frac{V'}{2T_{cl}}, \nonumber
\\  A =  \frac{\partial}{\partial x} + \frac{V'}{2T_{cl}}.
\end{eqnarray}
In the higher-dimensional many body problem, the gradients are relative to each of the Cartesian coordinates of all of the particles and $V$ is replaced by $V_{\sf N}(\{\vec{x}\})$.
Clearly ${\gamma} A^{\dagger} A \ge 0$ and thus if a zero-energy eigenstate of $H$ can be found it is the ground state. 
Now, the square root of the classical equilibrium distribution function, i.e., the wavefunction of Eq. (\ref{psie}) is clearly 
a null eigenstate of the operator $A$ above. Inserting Eq. (\ref{saa}) into Eq. (\ref{ssy}) leads to the identification of the quantum 
many body potential in terms of the corresponding classical potential energy $V_{\sf N}(\{\vec{x}\})$.
The astute reader will, up to trivial alterations, recognize these operators as the standard raising and lowering operators of the harmonic problem when $V$ is a harmonic potential. We 
briefly return to this point in subsection \ref{hs}. Basic general relation between quantum and classical systems for wavefunctions of the eikonal type are further discussed in Appendix \ref{sec:action}.
We provide very simple illustrative examples of the duality in Appendix \ref{ssimple}.

\subsection{Derivation of the quantum to classical correspondence for general dynamical correlation functions}
\label{TwoOperators}

The central role of this subsection is the derivation of Eqs.(\ref{qcqc},\ref{ggqc})
(or, more precisely, the derivation of Eqs. (\ref{gclassical}, \ref{gQuantum}) that will lead to Eqs. (\ref{qcqc},\ref{ggqc}).
The sole assumption made in the below derivation of Eqs. (\ref{gclassical}, \ref{gQuantum}) is that the classical system starts
from its equilibrium state and then evolves with some general (time dependent) potential $V_{\sf N}(t)$.
This will be mapped onto analytic continuations of the correlation and response functions
of a quantum system that starts at time $t=0$ in its ground state of Eq. (\ref{psie}) and then
evolved with the corresponding (time dependent) Hamiltonian $H(t)$. It is important
to emphasize that we make no assumptions regarding the final (and intermediate) states.
The classical (quantum) system need not stay in equilibrium (or within a ground state) as it evolves in time. 
Before detailing the derivation, we collect the basic relations discussed in subsection \ref{FPreview} with several new definitions,
\begin{eqnarray}
P(\{x\},t) =  \langle \{x\}|P(t)\rangle,
\label{ProbDist}
\\
P_{G} (\{x\},t) = \langle \{x\}|G \rangle = \frac{e^{- V_{\sf N}(\{x\})/T_{cl}}}{Z_{N}},
\label{GroundStateDist}
\\
H_{FP} |G\rangle = 0,
\label{GroundStateDef}
\\
\langle + | \{x\} \rangle = 1,
\label{UniformState}
\\
H = e^{ V_{\sf N}/2T_{cl}}H_{FP}e^{- V_{\sf N}/2T_{cl}}.
\label{QuantumClassicalHRelation}
\end{eqnarray}
These will serve as a point of departure for the calculations in this subsection. 
Eq. (\ref{ProbDist}) represents a general probability distribution in bra-ket notation. Eq. (\ref{GroundStateDist})
defines the ground state distribution as a Boltzmann distribution in bra-ket notation.  Eq. (\ref{GroundStateDef})
defines the ground state as the eigenvector of the Fokker-Planck Hamiltonian with zero eigenvalue.  
Eq. (\ref{UniformState}) defines the state $|+\rangle$ to be the uniform state such that $|+\rangle = \int d \{x\} |+\rangle$.
Lastly, Eq. (\ref{QuantumClassicalHRelation}) can be used to find a relationship between $H_{FP}$ and $H_{FP}^{\dagger}$.

Armed with these, we now proceed to some simple calculations. As $H$ is Hermitian,
\begin{eqnarray}
H^{\dagger} = e^{- V_{\sf N}/2T_{cl}}H_{FP}^{\dagger}e^{ V_{\sf N}/2T_{cl}}  \nonumber \\
(= H) = e^{ V_{\sf N}/2T_{cl}}H_{FP}e^{- V_{\sf N}/2T_{cl}}.
\end{eqnarray}
Explicitly multiplying by $e^{ V_{\sf N}/2T_{cl}}$ on the left and by $e^{- V_{\sf N}/2T_{cl}}$ on the right leads to
\begin{eqnarray}
H_{FP}^{\dagger} = e^{ V_{\sf N}/T_{cl}}H_{FP}e^{- V_{\sf N}/T_{cl}}.
\label{FPdaggerRelationFP}
\end{eqnarray}
We will now prove that the state $|+\rangle$ is a left eigenstate of the
Fokker-Planck Hamiltonian with zero eigenvalue.  Beginning with a simple extension of the definition of the ground
state,
\begin{eqnarray}
H_{FP} |G \rangle = 0 \rightarrow \langle G|H_{FP}^{\dagger} = 0.
\end{eqnarray}
As is evident from Eq. (\ref{FPdaggerRelationFP}), this is equivalent to
\begin{eqnarray}
\langle G|e^{ V_{\sf N}/T_{cl}}H_{FP}e^{- V_{\sf N}/T_{cl}} = 0,
\end{eqnarray}
which (from Eqs.(\ref{GroundStateDist}, \ref{UniformState})) implies that
\begin{eqnarray}
Z_{N}^{-1} \langle + | H_{FP}e^{- V_{\sf N}/T_{cl}} = 0.
\label{PlusEigenvector}
\end{eqnarray}
This illustrates that this uniform state is a left null eigenstate \cite{Biroli} for {\it all} Fokker-Planck Hamiltonians (i.e., $\langle +| H_{FP} =0$).

We will now derive our new central result of Eq. (\ref{qcqc}). Towards this end, we write anew the classical correlation function of Eq. (\ref{gclq}),
\begin{eqnarray}
\label{clear}
G_{\sf classical}(t)  = \langle \mathcal{O}_{1}(t) \mathcal{O}_{2} (0) \rangle.
\end{eqnarray}

By Bayes' theorem, the joint probability distribution, $P(\{\vec{x}\},\{\vec{y}\}) = P(\{\vec{x}\} | \{\vec{y}\}) P(\{\vec{y}\})$, the probability of finding coordinates $\{\vec{x}\}$ at time $t$ and coordinates $\{\vec{y}\}$ 
at time $0$, is given by the product of the conditional probability of finding $\{\vec{x}\}$ at  time $t$ given $\{\vec{y}\}$ at time $0$ with the probability of attaining $\{\vec{y}\}$ 
at time $t=0$.  For a lattice system with fields $\phi$ at different lattice sites (which we will briefly return to in Appendix \ref{aq2c}, the equality $P(\{\vec{x}\},\{\vec{y}\}) = P(\{\vec{x}\} | \{\vec{y}\}) P(\{\vec{y}\})$ is to be replaced by 
$P(\{\phi(\vec{x},t)\}, \{\phi(\vec{x},0)\}) = 
P(\{\phi(\vec{x},t)\}| \{\phi(\vec{x},0)\}) 
P( \{\phi(\vec{x},0)\})$. As discussed in Appendix \ref{FPreview},
the ground state has a probability distribution given 
by a Boltzmann distribution  $Z_{N}^{-1} e^{-\beta V_{\sf N}
(\{y\})}$ [see Eq. (\ref{GroundStateDist})].  The conditional probability $P(\{x\} | \{y\})$ can be expressed in terms of the 
matrix element of ${\cal{T}} e^{-\int_{0}^{t} H_{FP}(t') dt'}$ (where $\cal{T}$ is the time ordering operator)
as this conditional $P$ satisfies Eq. (\ref{FPEQ}).  With $\mathcal{O}_{1}$ depending on the coordinates $\{\vec{y}\}$ at time $t$ 
and $\mathcal{O}_{2}$ on the coordinates $\{\vec{x}\}$ at time $t=0$, all 
this implies the form of the expectation value of Eq. (\ref{clear}),
\begin{eqnarray}
\label{fcllong}
G_{\sf classical}(t)  = \int d^{dN}x~ d^{dN}y ~\mathcal{O}_{1} P(\{\vec{x}\},\{\vec{y}\}) \mathcal{O}_{2} \nonumber
 \\ = \int d^{dN}x~ d^{dN}y ~\mathcal{O}_{1} P(\{\vec{x}\}|\{\vec{y}\}) \mathcal{O}_{2} P(\{\vec{y}\}) \nonumber
 \\ = \int d^{dN} x~ d^{dN} y~ \mathcal{O}_{1} \langle \{\vec{x}\} | {\cal{T}} e^{-\int_{0}^{t} H_{FP}(t') dt'}|\{\vec{y}\} \rangle 
  \mathcal{O}_{2}\nonumber \\  \times \frac{e^{-\beta V_{\sf N}(\{\vec{y}\})}}{Z_{N}}. 
  %\nonumber
\end{eqnarray}
It is important to re-iterate and emphasize yet again that, in the last line above, we only assume that {\it the initial state} ($|\{\vec{y}\} \rangle$ at time $t=0$) {\it is
in thermal equilibrium}. {\it The system need not be in thermal equilibrium at positive times.} As stated earlier, this is the sole assumption made in this derivation
for general time dependent systems with dynamical $V_{\sf N}$ (and thus for time dependent Fokker-Planck 
operators). A similar derivation would hold {\it mutatis  mutandis} when the system is initially out of equilibrium 
and is in equilibrium in its final state. 

Eq. (\ref{UniformState}) asserts 
 that $\int d^{dN}x \langle \{ \vec{x}\}| = \langle +|$. Invoking this along with Eq. (\ref{GroundStateDist}), we have that
\begin{eqnarray}
\label{fcla}
G_{\sf classical}(t)  = \langle + | \mathcal{O}_{1}  {\cal{T}}e^{-\int_{0}^{t} H_{FP}(t') dt'} \mathcal{O}_{2} |G \rangle.
\end{eqnarray}
As is evident from Eq. (\ref{PlusEigenvector}), inserting an exponentiation of $H_{FP}$ to the right
of the state $\langle + |$ leads to a multiplication by unity. Thus Eq. (\ref{fcla}) can be rewritten as
\begin{eqnarray}
G_{\sf classical}(t) = \langle + |{\cal{T}} e^{\int_{0}^{t} H_{FP}(t') dt'} \mathcal{O}_{1} \nonumber \\ \times {\cal{T}} e^{-\int_{0}^{t} H_{FP}(t') dt'} \mathcal{O}_{2} |G\rangle.
\end{eqnarray}
With the aid of Eq. (\ref{QuantumClassicalHRelation}), we can express this quantity in terms of the quantum Hamiltonian $H$ instead of $H_{FP}$, 
\begin{eqnarray}
 G_{\sf classical}(t)  = \langle + |e^{-V_{\sf N}/(2T_{cl})} {\cal{T}} e^{\int_{0}^{t} H(t') dt'} \mathcal{O}_{1} \nonumber \\ \times {\cal{T}}  e^{-\int_{0}^{t} H(t') dt'} e^{V_{\sf N}/(2T_{cl})}\mathcal{O}_{2} |G\rangle.
\end{eqnarray}
Rather explicitly multiplying and dividing by $\sqrt{Z_{N}}$,
\begin{eqnarray}
\label{fcle}
G_{\sf classical}(t)  = \langle + |\frac{e^{-
 V_{\sf N}/(2T_{cl})}}{\sqrt{Z_{N}}}  &{\cal{T}}& e^{\int_{0}^{t} H(t') dt'} \mathcal{O}_{1} {\cal{T}} e^{-\int_{0}^{t} H(t') dt'} \nonumber \\ &\times& \mathcal{O}_{2} 
 \sqrt{Z_{N}} e^{V_{\sf N}/(2T_{cl})} |G\rangle.
\end{eqnarray}
As discussed in subsection \ref{FPqs}
(in particular, Eq. (\ref{psie})),
the ground state of the quantum system is given by $|0\rangle = \sqrt{Z_{N}} e^{V_{\sf N}/(2T_{cl})} | G \rangle$.
Further invoking Eqs. (\ref{GroundStateDist},\ref{UniformState}), we can 
rewrite Eq. (\ref{fcle}) as
\begin{eqnarray}
\label{gclassical}
G_{\sf classical}(t) = \langle 0| {\cal{T}} e^{\int_{0}^{t} H(t') dt'} \mathcal{O}_{1}  \nonumber
\\ \times {\cal{T}} e^{-\int_{0}^{t} H(t') dt'} \mathcal{O}_{2} |0 \rangle.
\end{eqnarray}
Note that, in this equation,
$|  0 \rangle$ is the ground state of the system defined by the quantum Hamiltonian $H$. 
Our results above are general. We will shortly use Eq. (\ref{gclassical}) in order to relate it to correlations in the quantum system. 
Under the exchange of $t$ by $(it)$, the reader may recognize Eq. (\ref{gclassical}) as a correlation function in the quantum system.
One very simple point which is worth emphasizing is that not only the ground state of Eq. (\ref{psie}) is trivially normalized but, of course,
any state formed by the evolution with the unitary time ordered exponential ${\cal{U}}(t) = e^{-i \int_{0}^{t} H(t') dt'}$.

In the quantum arena, it is clear that for a system initially prepared in the ground state $|0 \rangle$
and then evolved with some Hamiltonian $H(t)$, the 
corresponding correlation function is given by 
\begin{eqnarray}
\label{gQuantum}
 G_{\sf Quantum} (t)= \langle 0| {\cal{T}} e^{i\int_{0}^{t} H(t') dt'} {\mathcal O}_{1}  &{\cal{T}}& e^{-i\int_{0}^{t} H(t') dt'} \nonumber
 \\ &\times&  {\mathcal O}_{2} | 0 \rangle.
\end{eqnarray}
By comparing Eq. (\ref{gclassical}) with Eq. (\ref{gQuantum}), Eq. (\ref{qcqc}) immediately follows. 
The fundamental relation of Eq. (\ref{qcqc}) establishes the connection between the overdamped Langevin equation of a classical particle at finite temperature
and the Schr\"odinger equation of the dual quantum Hamiltonian. We suspect that related to this result is the fluctuation-dissipation theorem that relates correlation functions with the expectation value of time-ordered products in equilibrium, see for example Ref.~\onlinecite{Parisi} (Chap.~13).

A derivation similar to that above can be performed for a correlation function involving an arbitrary number of operators.
In the classical arena, such a correlation function takes the form of
\begin{eqnarray}
G_{classical} = \langle \mathcal{O}_{1}(t_{1})\mathcal{O}_{2}(t_{2})...\mathcal{O}_{n}(t_{n}) \rangle,
\end{eqnarray}
where $\mathcal{O}_{i}$ are arbitrary operators and $t_{1} < t_{2} < ... < t_{n}$.
Similar to our earlier calculations, by Bayes' theorem, this correlation function is given by
\begin{eqnarray}
\nonumber \int d ^{dN}x_{1}  ~ d^{dN}x_{2} ~  ...d^{dN} x_{n} ~ \mathcal{O}_{n}\nonumber
\\ \times  \langle \{\vec{x}_{n}\} | {\cal{T}}e^{-\int^{t_{n}}_{t_{n-1}}H_{FP}(t')dt'}|\{\vec{x}_{n-1}\} \rangle 
\mathcal{O}_{n-1} \nonumber
\\ ... \langle \{\vec{x}_{2}\} | 
{\cal{T}}e^{-\int^{t_{2}}_{t_{1}}H_{FP}(t') dt'}|\{\vec{x}_{1}\} \rangle~ \mathcal{O}_{1}~ \frac{e^{-\beta V_{N}(\{\vec{x}_{1}\})}}{Z_{N}}. 
%\nonumber
\end{eqnarray}
Invoking identity matrix insertions and integrations over a complete set of eigenstates as before, this reduces to
\begin{eqnarray}
 \langle + |{\cal{T}}e^{-\int^{t_{1}}_{t_{n}}H_{FP}(t')dt'} \mathcal{O}_{n} {\cal{T}}e^{-\int^{t_{n}}_{t_{n-1}}H_{FP}(t')dt'}\mathcal{O}_{n-1} \nonumber \\ 
... {\cal{T}}e^{-\int^{t_{2}}_{t_{1}}H_{FP}(t')dt'} \mathcal{O}_{1} |G \rangle.
\end{eqnarray}
Transforming to the quantum Hamiltonian $H(t)$ and its respective ground state at time $t=0$ yields
\begin{eqnarray}
 \langle 0 | {\cal{T}}e^{-\int^{t_{1}}_{t_{n}}H(t')dt'} \mathcal{O}_{n} {\cal{T}}e^{-\int^{t_{n}}_{t_{n-1}}H(t')dt'}\mathcal{O}_{n-1} \nonumber \\ ... {\cal{T}}e^{-\int^{t_{2}}_{t_{1}}H(t')dt'} \mathcal{O}_{1}|0 \rangle.
\end{eqnarray}
The remainder of the derivation is similar to that in the two time case.
In order to transition from the classical to the quantum system, we replace, in all pertinent correlation functions, the times $\{t_{a}\}_{a=1}^{n}$ by $\{it_{a}\}_{a=1}^{n}$.

We next return to the two time correlation function and discuss the quantum response function $R_{\sf Quantum}$
that monitors the change in the average value of ${\mathcal O}_{1}$ as a result of a perturbation ${\mathcal O}_{2}$.
We first review standard textbook \cite{forster} results concerning quantum linear response 
functions and then invoke our new result of Eq. (\ref{qcqc}) to obtain zero
temperature quantum linear response functions given corresponding results on finite temperature classical duals. 
Towards this end, we first consider the Hamiltonian
\begin{eqnarray}
H_{\sf tot} = H + H^{'},
\end{eqnarray}
where $H'$ is a small perturbation which can be expressed as $H'=- \lambda \mathcal{O}_{2}$. 
We next review the standard protocol for computing the lowest order deviation
\begin{eqnarray}
\delta \langle {\mathcal O}_{1} (t) \rangle = \langle {\mathcal O}_{1} (t) \rangle_{\lambda} -  \langle {\mathcal O}_{1} (t) \rangle_{0},
\end{eqnarray}
which we will evaluate within the ground state $| 0 \rangle$. 
This deviation is readily computed within the interaction picture where we evolve with the time ordered exponential $T \exp(-i H' t)$,
\begin{eqnarray}
 \langle {\mathcal O}_{1} (t) \rangle_{\lambda}  \approx \langle \left(1- i \int^{t} dt' ~ \lambda(t)~ {\mathcal O}_{2}(t')\right) {\mathcal O}_{1}(t)  \nonumber \\ \times
  \left(1+ i \int^{t} dt' ~\lambda(t)~ {\mathcal O}_{2}(t')\right) \rangle.
 \end{eqnarray}
Collecting terms to lowest order,
\begin{eqnarray}
\delta \langle \mathcal{O}_{1}(t)  \rangle \approx i \int^{t} dt' ~ \lambda(t') ~ \langle [\mathcal{O}_{1}(t),\mathcal{O}_{2}(t')] \rangle \nonumber
\\ = i \int_{0}^{\infty} d \tau ~ \lambda (t-\tau)~ \langle [\mathcal{O}_{1}(\tau),\mathcal{O}_{2}(0)] \rangle \nonumber
\\ \equiv \int_{-\infty}^{\infty}  d \tau ~\lambda(t-\tau)~ R_{\sf Quantum}(\tau).
\label{LinearInFofT}
\end{eqnarray}
As ${\mathcal{O}}_{1}(t) = {\cal{T}} e^{-i\int_{0}^{t} H(t') dt'} {\mathcal{O}}_{1} {\cal{T}} e^{i\int_{0}^{t} H(t') dt'}$, from the last line of Eq. (\ref{LinearInFofT}), the quantum response function

\begin{eqnarray}
\label{rquantum}
R_{\sf Quantum}(t) = i \Theta(t) \langle 0 | \big[ {\cal{T}} e^{i\int_{0}^{t} H(t') dt'} \mathcal{O}_{1}  \nonumber \\ \times {\cal{T}} e^{-i\int_{0}^{t} H(t') dt'},\mathcal{O}_{2} \big] | 0 \rangle.
\end{eqnarray}
Comparing Eqs. (\ref{gclassical}, \ref{rquantum}), we derive Eq.(\ref{ggqc})
%which we rewrite here for ease 
by further expanding it to get the imaginary part of the analytically continued classical 
correlation function,
\begin{eqnarray}
R_{\sf Quantum}(t) &=& i  \Theta(t) (G_{\sf classical}(it) - G^{*}_{\sf classical}(it)) \nonumber
\\ &=& - 2 \Theta(t)  \Im\,  G_{\sf classical}(it).
\label{RQuantumVsGClassical}
\end{eqnarray}

\subsection{Fields on a lattice}
\label{latfi}
We conclude this section with a brief discussion of the duality for fields on lattice sites.
If we replace Eq. (\ref{Langevin})
by 
\begin{equation}
  \gamma_{i} \frac{d\phi_{i}}{dt} = - \frac{\delta}{\delta \phi_{i}} 
                                 V_{\sf N}(\phi_{1}, ..., \phi_{N}) + {\eta}_{i}(t),
  \label{Langevin'}
\end{equation}
to describe a classical lattice system with fields $\phi_{i}$ at the various lattice points $i$,
then trivially replicating all of our calculations thus far with the exchange $\vec{x}_{i} \to \phi_{i}$ (including in all gradient or variational derivative operators), we will arrive at a corresponding quantum 
lattice system {\it mutatis mutandis}.

\section{High Dimensional Quantum Glass Models Derived from Classical counterparts}
\label{models}
In the next sections, we will examine the consequences of Eq. (\ref{qcqc}) for disparate quantum systems. As outlined earlier, our basic three-prong approach will be rather simple: \newline
{\bf{(1)}} We take a classical dissipative system whose dynamical behavior is known at finite temperatures (including, in particular, pertinent temporal correlation functions of the form of Eq.(\ref{gclq})). \newline
{\bf{(2)}} We determine the corresponding dual quantum Hamiltonian using Eq. (\ref{HQM}) or its explicit form for classical pair potentials
which we detail below. \newline
{\bf{(3)}} We next invoke Eq. (\ref{qcqc}) in order to determine the very same correlation function of Eq. (\ref{gclq}), yet now at zero temperature for the dual quantum Hamiltonians.  \newline

In this section, we will detail a few (of the many known) classical glassy systems for which our results for the quantum duals will hold. 
The specific heavily investigated classical dissipative systems that we focus on all exhibit glassy dynamics. It is worth emphasizing that the results that 
we will obtain using our three-prong approach {\it do not rely on any special integrability of the quantum models}. Rather, by using
the multitude of available information on the quantum glass system, we will be able to make exact statements on numerous quantum
systems. 

We consider what specifically 
occurs when the classical potential energy in Eq. (\ref{Langevin}) is the sum of pairwise interactions (as it typically is), 
\begin{eqnarray}
  V_{\sf N}(\{\vec{x}\})  = \frac{1}{2} \sum_{i \neq j} V_{ij}(\vec{x}_{i} 
                       - \vec{x}_{j}).
                       \label{vnc}
\end{eqnarray}
For such systems, the quantum many body Hamiltonian of Eq. (\ref{HQM}) explicitly contains
an effective potential which is the sum of two and three body interactions,
\begin{eqnarray}
&&{\cal{V}}_{\sf Quantum}(\{\vec{x}\})  =  \sum_{i} \frac{1}{\gamma_{i}} \left[-\frac{1}{2} \nabla_{i}^{2} V_{\sf N}
                                 + \frac{1}{4T_{cl}} (\vec{\nabla}_{i} V_{\sf N})^{2}\right]  \nonumber\\ 
                          &&=  - \frac{1}{2} \sum_{i \neq j}  \frac{1}{\gamma_{i}} \nabla_{i}^{2} 
                                 V_{ij} %\nonumber\\
                               +  \sum_{i;j \neq i; j' \neq i} \frac{\vec{\nabla}_{i} V_{ij}   
\cdot\vec{\nabla}_{i} V_{ij'}}{4T_{cl}\gamma_{i}}.
                                \label{vQ}
\end{eqnarray}

For a given classical two body potential in $d$ dimensions which is both translationally and rotationally invariant, $V(\vec{x}) = V(|\vec{x}|)$, the resulting
quantum potential energy is given by (as in \cite{Biroli} yet now trivially extended to general classical temperatures $T_{cl}$),
\begin{eqnarray}
{\cal{V}}_{\sf Quantum}(\{\vec{x}\}) &=& \frac{1}{2} \sum_{i \neq j} v_{\sf Quantum}^{pair}(\vec{x}_{i} - \vec{x}_{j}) \nonumber\\ 
  & & + \sum_{i; j \neq i; j' \neq i} v^{3-body}_{\sf Quantum} (\vec{x}_{i}-\vec{x}_{j}, \vec{x}_{i} - \vec{x}_{j'}); \nonumber\\
      v^{pair}(\vec{x}) &=& - \nabla^{2} V(\vec{x}) + \frac{1}{2T_{cl}} [\vec{\nabla} V(\vec{x})]^{2} \nonumber\\ 
  &=& - \frac{d-1}{r} V^{\prime}(r) - V^{\prime \prime}(r) + \frac{1}{2T_{cl}} [V^{\prime}(r)]^{2}; \nonumber\\
      v^{3-body}(x, x') &= &\frac{1}{4T_{cl}} \vec{\nabla} V(x) \cdot \vec{\nabla} V(\vec{x}') \nonumber\\ 
  &=& \frac{1}{4T_{cl}} \frac{\vec{x}}{r} \cdot \frac{\vec{x}'}{r'} V'(r) V'(r'),
  \label{Potential}
\end{eqnarray}
with $r=|\vec{x}|= |\vec{x}_{i}-\vec{x}_{j}|$ and, in the three-body term, $r'=|\vec{x}'|= |\vec{x}_{i} - \vec{x}_{j'}|$. 
For short range classical interactions $V(r)$, the three-body term can be appreciable only if the 
three points $(i,j,j')$ in the second sum of Eq. (\ref{Potential}) defining the distances $r$ and $r'$ all lie within the short distance of one another where the classical potential operates. Thus, statistically,
the three-body interactions are typically insignificant by comparison to,
the far more dominant, pair interaction terms in Eq. (\ref{Potential}). 

As we explained in section \ref{FPreview}, whenever the classical potential $ V_{\sf N}$ is invariant under the permutations of the coordinates of any pair 
of particles (as it explicitly is when it is the sum of symmetric pair interactions $V(|\vec{x}_{i} - \vec{x}_{j}|)$) then the resulting
quantum many body system is bosonic. In Appendix \ref{sslater} (and, to a lesser degree in Appendix \ref{aq2c}), we elaborate how our results can, as a matter of principle, be explicitly extended
to specific fermionic systems (which are of great pertinence in our goal of illustrating the feasibility of the behaviors that we study in this work for 
electronic systems). 

Although our results in the sections that follow are very general, it is nevertheless useful to have concrete models in mind. 
We next detail some typical model systems that we will refer to. These systems include both off-lattice liquid and lattice systems. The liquids that we list exhibit fluid behavior at high classical temperatures and glass like features at high densities and/or low temperatures. {\it Within the highly viscous low temperature regime, the classical fluids that we list below become overdamped and may be modeled by Eq. (\ref{Langevin})}.  \newline

$\bullet$ {\underline{{\it{Liquid models:}}}}\newline
({\bf{A}}) As a first example we list a system of three-dimensional spheres ($s$).The classical potential associated with this system is given by $V_{s}(r) = V_{0} \exp(-\lambda [(r/\sigma)^{2}-1])$. This model has been extensively studied \cite{Biroli}. In this system, the potential $V_{s}$ has a clear finite
range (set by the diameter of the spheres $\sigma$). Following our earlier discussion, 
the magnitude of the three-body term in Eq. (\ref{Potential})
will be negligible by comparison to that of the pair interactions. The corresponding pair term set by Eq. (\ref{Potential}) is 
\begin{eqnarray}
\label{pairq}
v^{pair}(r) = \frac{2 \lambda d - 4 \lambda^{2} r^{2}}{\sigma^{2}} V_{s}(r) + \frac{2 \lambda^{2} r^{2}}{T_{cl} \sigma^{4}} [V_{s}(r)]^{2}.
\end{eqnarray}
In the limit $\lambda \to \infty$, the classical system corresponds to that of hard spheres
where $\sigma$ is the diameter of the hard sphere and the quantum potential of Eq. (\ref{pairq})
similarly exhibits a dominant hard sphere repulsion (augmented by an attractive potential
at the sphere boundaries that is of range $1/\lambda$). In the sections that follow we will
refer to the finite temperature ($T_{cl}>0$) behavior of this system (and the other
models below). In the hard sphere ($\lambda \to \infty$) limit, 
this system becomes temperature independent.

({\bf{B}}) A classical bi-disperse repulsive system given by the pair potential \cite{repulsive}
\begin{eqnarray}
\label{bi-disperse}
V_{ab}(r) = \epsilon \Big( \frac{\sigma_{ab}}{r} \Big)^{12}
\end{eqnarray} 
between two particles $(a,b)$ of two possible types
($(a,b) \in 1,2$) with $\sigma_{ab} = (\sigma_{a} + \sigma_{b})/2$ and $\sigma_{2}/\sigma_{1} = 1.2$.
The corresponding quantum potential is given by Eq. (\ref{Potential}). It is this full potential
that leads to the exact same dynamical correlation functions for the quantum system following Eq. (\ref{qcqc}).
Similar to (A), for pair distances larger than $\sigma_{ab}$, the 3-body term in the quantum Hamiltonian of Eq. (\ref{Potential}) is far smaller than the pair interaction term.
Thus, at low temperatures, $T_{cl} \ll \epsilon$, in any number of dimensions $d$, this classical system has a quantum dual given by
a pair potential 
\begin{eqnarray}
v^{pair}_{ab}(r) \simeq \frac{72 \epsilon \sigma_{ab}^{24}}{ T_{cl}  r^{26}}.
\end{eqnarray}

({\bf{C}}) A classical bi-disperse Lennard-Jones mixture.
Similar to (B), this is a model of two species: 1,2. Unlike (B). however, 
each pair of atoms $(a,b)$ interact via a Lennard-Jones type interaction,
\begin{eqnarray}
\label{KA}
V_{ab}(r) =  \epsilon_{ab} \Big[ \Big( \frac{\sigma_{ab}}{r} \Big)^{12} - \Big( \frac{\sigma_{ab}}{r} \Big)^{6} \Big].
\end{eqnarray}
This augments the repulsive only potential of (B) by an additional longer range attractive interaction. 
In the standard Kob-Andersen mixtures \cite{KA} that we will refer to,  $\epsilon_{12}/\epsilon_{11} = 1.5, \epsilon_{22}/\epsilon_{11} = 0.5$
with similar lengths $\sigma_{ab}$ as in (B).  The zero temperature quantum dual of a low temperature classical system ($T_{cl} \ll \epsilon_{11}$) is
\begin{eqnarray}
v^{pair}_{ab}(r) \simeq \frac{18 \epsilon_{ab}^{2} \sigma_{ab}^{12}}{T_{cl} r^{14}} \Big[ 1 - 4 \Big\{ \Big( \frac{\sigma_{ab}}{r} \Big)^{6} \nonumber
\\ - \Big( \frac{\sigma_{ab}}{r} \Big)^{12} \Big\} \Big].
\end{eqnarray}

$\bullet$ {\underline{{\it Lattice models:}}\newline
({\bf{D}}) The N3 and N2 lattice models \cite{2DN3,3DN2} (which share some similarity with earlier lattice glass models \cite{BM}) .  In the square lattice N3 model, particles are endowed with hard core repulsive interactions that
extend up to a distance of three steps on the lattice. Similarly, in the cubic lattice N2 model particles cannot be nearest neighbors nor next nearest neighbors
(i.e., the repulsive hard-core interactions extend up to a distance of two steps on the lattice). In sections \ref{ls}, \ref{esp}, we will further motivate and discuss
quantum lattice systems.

At their core, the results that we discuss next are not limited to the examples (A-D) above nor to simple classical pair interactions. 
Given any classical system whose evolution is given by Eqs. (\ref{Langevin}, \ref{FPEQ}) the corresponding dual quantum system is provided by
Eqs. (\ref{HQM},\ref{QMAC}).
This can, e.g., include models of classical dislocation motion and turbulence in liquids.

\section{Glassy dynamics in off-lattice quantum fluids}
\label{glassy}
Armed with all of the background and results described in earlier sections, we now proceed to derive general physical results in quantum systems.
Our aim is to show that as a matter of principle zero-point quantum fluctuations can lead to very rich glass type behaviors in numerous many body systems
which mirror those that appear in dissipative classical systems at finite temperatures. As we alluded to in the Introduction, classical liquids may become quenched into a glassy
state when they are rapidly cooled (``supercooled'') below structural freezing temperatures and fall out of thermal equilibrium. Invoking Eq. (\ref{qcqc}), this will suggest that in the zero-temperature quantum duals a corresponding  
phenomenon will occur $-$ quantum fluids may veer towards a glassy state as parameters are rapidly changed
in time. As we emphasized earlier, our derivation of subsection \ref{TwoOperators} allowed
(as is physically crucial) for time dependent Hamiltonians which emulate rapid
changes in the classical temperature or any other parameters in the interaction
and for classical final (or initial) states which are out of equilibrium. 
Its sole assumption was that the average over the noise at
any instant of time was still afforded by Eq. (\ref{noiseeq}) with
$T_{cl}$ the corresponding classical temperature at
that time. We focus on measurable quantities that may be ascertained
from response functions.

Response functions in classical glass forming liquids which become
progressively more viscous and become frozen into a glass as their temperature is rapidly lowered (as well as response functions in various electronic systems), 
suggest the presence of a distribution of local relaxation times that lead to, e.g., the canonical Cole-Cole or Cole-Davidson \cite{Phase1, Phase2} 
and similar forms as we briefly elaborate on. In various guises, all of the models discussed in section \ref{models} exhibit glass like features
including notably the distribution of relaxation times which we discuss now.

The response of a single attenuated mode to an initial impulse at time $t=0$ 
scales as
$g_{single} \sim  \exp(-t/\tau)$ with $\tau$ the relaxation time; the Fourier transform of this response is
$g_{single}(\omega) = g_{0}/(1-i \omega \tau)$. 
In systems with a {\it distribution} $f(\tau')$ of relaxation events, the response functions are given by
$\int d\tau' f(\tau') \exp(-t/\tau')$. Empirically, in dissipative plastic or visco-elastic systems, relaxations scale as $\exp[-(t/\tau)^{c}]$ with a power $0<c<1$ that 
leads to a ``stretching'' of the response function.  
This stretched exponential and other similar forms, such as the Cole-Cole (CC) and Davidson-Cole (DC) functions, quintessentially capture the distribution of
relaxation times. \cite{Phase1, Phase2} With $g(\omega) = g_{0} G(\omega)$,
where $g_{0}$ is a constant,
the CC and CD forms correspond to different choices of $G$,
\begin{eqnarray}
\label{CCD}
G_{CC}(\omega)  = \frac{1}{[ 1- (i \omega \tau)^{\overline{\mu}}]}, \nonumber
\\ G_{DC}(\omega)  = \frac{1}{[ 1- i \omega \tau]^{\overline{nu}}}.
\end{eqnarray} 
The parameters $\overline{\mu}$ and $\overline{\nu}$ qualitatively emulate the real-time stretching exponent $c$. 
This distribution of relaxation times might be associated with different local dynamics (dynamical heterogeneities) to
which we will turn to shortly in subsection \ref{qdhsub}.  As liquids are supercooled, their characteristic relaxation times and viscosity may increase dramatically.
There are several time scales that govern the dynamics of supercooled liquids. 
The so-called ``$\alpha$ (or primary) relaxation'' is associated with cooperative motion and leads to a pronounced rise of the viscosity (especially so in the ``fragile'' glass-formers). 
Empirically, in real classical supercooled liquids at a temperature $T_{cl}$, the $\alpha$ relaxation times follow the Vogel-Fulcher-Tammann form \cite{Rault00},
\begin{eqnarray}
\label{VFT_eq}
\tau(T_{cl}) =
\left\{
\begin{array}{ll}
\tau_0 \,e^{D T_{0}/(T_{cl}-T_0)} & \mbox{ for $T>T_0$} , \\
 \infty  & \mbox{ for $T \le T_0$} .
\end{array}
\right.
\end{eqnarray}
Here, $T_{0}$ is the temperature of the classical system at which the relaxation times (if Eq. (\ref{VFT_eq}) is precise) will 
truly diverge and $D$ is a dimensionless constant. Mode coupling theory \cite{MCT} and numerous
other theories might similarly capture aspects of the increase in the $\alpha$ relaxation time.  
In a low temperature liquid, augmenting the long time $\alpha$ relaxations to equilibrium, there are so-called ``$\beta$ (or secondary) relaxations'' \cite{Johari}
which further manifest in local relaxation processes.  The $\beta$ relaxation times scale with temperature in an Arrhenius type fashion, 
\begin{eqnarray}
\label{secondary}
\tau_{\sf secondary} \sim e^{\Delta/ T_{cl}}
\end{eqnarray}
with $\Delta$ a temperature independent constant.
Recent work suggests intriguing relations between $\alpha$ and $\beta$ relaxations \cite{alphabeta}.  
By virtue of Eq. (\ref{qcqc}), the finite temperature classical $\alpha$ (and $\beta$) relaxations
and their associated stretched exponential type relaxations all have zero temperature quantum duals.
In the quantum arena, as discussed in Section \ref{rev-sec}, the classical temperature $T_{cl}$ is replaced by the effective mass $m= \gamma/(2T_{cl})$
and parametrical changes in the many body potential ${\cal{V}}_{\sf Quantum}(\{\vec{x}\})$. 

The classical hard sphere limits of model A and model D of Section \ref{models} are athermal; in these models a glass type state may only be arrived at by varying the system density.
By contrast, in models B and C (as well as away from the hard sphere limit of model (A)) lowering the temperature may induce a transition into a glass.  
By a trivial application of our result of Eq. (\ref{ggqc}), all of these finite temperature classical forms have the same quantum zero-temperature counterparts. 
Thus, {\it  the relaxation times in the quantum dual models scale in precisely the same way as they do in the classical glass forming systems} (including Eq. (\ref{VFT_eq}) for 
duals to classical glass formers). 
Slower dynamics also appears as the system density increases. Several works, e.g. \cite{density_s}, suggest that relaxation times are a function
of a composite quantity involving both density and temperature. It should be noted that all of our derivations start from Eq. (\ref{Langevin}).
When examining the quantum dual to empirical forms describing classical liquids, 
the bare viscosity (or associated bare relaxation time $\tau_0$ in Eq. (\ref{VFT_eq})) of the ambient liquid appearing in these equations of motion may, in principle, be allowed to change as the temperature (and density) 
are varied. These may appear in addition to changes in $T_{cl}$ and $V_{\sf N}$ (capturing, e.g., changes in the density).
In classical simulated liquids, the bare viscosity may be kept constant.

\subsection{Quantum dynamical heterogeneities and relations for four-point correlators}
\label{qdhsub}

We now focus on an intriguing aspect of classical glasses which by virtue of the relation of Eqs. (\ref{qcqc},\ref{ggqc})  [as alluded to 
in \cite{physics_ZN}] leads to the appearance of new dynamical correlations in quantum systems. 
Disorder free models for classical glass formers (including various simulated quenched systems such as those endowed with 
various classical potentials $V_{N}$ discussed in section \ref{models} that do not permit simple crystalline orders) are known to 
exhibit ``dynamical heterogeneities" (DH),
i.e., a non-uniform distribution of local velocities \cite{het.}
with the location of the more rapidly moving particles changing with time. By invoking Eqs. (\ref{qcqc},\ref{ggqc}), we see that {\em Quantum Dynamical
Heterogeneities} (QDH) \cite{physics_ZN} appear in their corresponding zero temperature quantum counterparts. That is, in disorder free quantum systems derived (via Eq. (\ref{HQM})) from the corresponding classical
systems, zero point dynamics is spatially non-uniform.

The presence of DH is empirically seen by numerous probes \cite{het.} in real glass formers as well as model systems (including all of the systems in section \ref{models}).
As we alluded to earlier, one often used metric is that of the four-point correlations of Eq. (\ref{g4xy}) in various guises. 
These correlation functions are of the form of Eq. (\ref{gclq}) with ${\cal{O}}(t)$ denoting the overlap between fields $\phi$ when these are separated in time,   
\begin{eqnarray}
{\cal{O}}_{\vec{q}} (t) = \phi_{\vec{q}}(t) \phi_{-\vec{q}}(0) - \langle \phi_{\vec{q}}(t) \rangle \langle \phi_{-\vec{q}}(0) \rangle,
\label{oqS}
\end{eqnarray}
with $\vec{q}$ any wave-vector. 
When Eq. (\ref{oqS}) is substituted into Eqs. (\ref{gclq},\ref{qcqc}), the four-point correlator can be computed. Classically, the Fourier space correlation functions (denoted $S^{\sf classical}_{4}(\vec{q},t)$ below) typically have an Ornstein-Zernicke type or similar related form, e.g., \cite{karmakar} 
\begin{eqnarray}
S^{\sf classical}_{4}(\vec{q},t) = \frac{\chi_{4}(t)}{1+q^{2} \xi_{4}(t)^{2}},
\label{Scl}
\end{eqnarray}
with the length scale $\xi_{4}(t)$ representing the size of the typical dynamical heterogeneity, when the system is examined at two times separated by an interval $t$.
The four-point susceptibility $\chi_4(t)$ is simply set by $\int d^{d}x \, G_4(\vec{x},t)$ (see Eq. (\ref{g4xy+0}).
The key feature of Eq. (\ref{Scl}) is that all of the $q$ dependence has been relegated to a Lorentzian form while $\chi_{4}$ and $\xi_{4}$ are otherwise general
time dependent functions. We may next invoke Eqs. (\ref{qcqc},\ref{ggqc}) to generate the quantum counter-part of Eq. (\ref{Scl}) (or of any other related form) and Fourier transform to real space
to obtain, in the notation of Eq. (\ref{g4xy}), the spatial quantum correlation function $G^{\sf Quantum}_{4}(\vec{x}-\vec{y},t)$ associated with the potential ${\cal{V}}_{\sf Quantum}$ of Eq.(\ref{Potential}).
The Fourier integral will be dominated by momentum space poles at $q=\pm i \xi^{-1}_{4}$. It is clear that in employing the transformation of Eqs. (\ref{qcqc},\ref{ggqc}),  $G^{\sf Quantum}_{4}(\vec{x}-\vec{y},t)$ 
will exhibit exponential decay with the very same correlation length $\xi_{4}$ that is present in the classical system.  This affords a direct proof of the dynamical length scale $\xi_{4}$ in
{\it all zero temperature quantum counter-parts} (given by Eq. (\ref{HQM})) to any dissipative classical system that is known to exhibit these (and there are numerous known classical systems that
exhibit dynamical heterogeneities \cite{het.}). 

Following our mapping, an exponentially decaying real space 4-point correlator ($r=|x-y|$) in the classical problem ($\exp(-r/\xi_{4})$) will
lead to an oscillatory decay in the quantum dual. Specifically, if the 4-point correlation length in the classical problem diverges as $\xi_{4} \sim \tau^{1/z}$ then,
as we derive in detail in Appendix \ref{dets}, the correlation function of Eq. (\ref{g4xy}) 
for the dual quantum system will be given by 
\begin{eqnarray}
G_{4, \sf{Quantum}}(r,t) && \sim \sin(\frac{r}{\tau^{1/z}} \sin \frac{\pi}{2 z}) 
 \nonumber
\\ && \times \exp[-\frac{r}{\tau^{1/z}} \cos \frac{\pi}{2z}].
\end{eqnarray}

\subsection{Rapidly increasing time scale with concomitant slowly increasing length scale in quantum glasses}

There is a proof that a growing static length scale must accompany 
the diverging relaxation time of glass transitions \cite{montanari}.
Some evidence has indeed been found for growing correlation lengths (static and those
describing dynamic inhomogeneities)
\cite{ref:mosayebiCLSGT,ref:berthierCL,ref:karmakarsastry}.
As we noted earlier in this work, 
correlation lengths were studied via ``point-to-set'' 
correlations \cite{BiroliPointToSet,ps} and pattern repetition size \cite{kl}.  Other current common methods of characterizing structures include (a) Voronoi polyhedra, \cite{ref:aharonov,sheng,ref:finney}, (b) Honeycutt-Andersen indices \cite{HA},  and (c) bond orientation \cite{BO}; 
all centering on an atom or a given link. More recent approaches include graph theoretical tools \cite{graph_method}. Not withstanding current progress, it is fair to say that currently 
most ``natural'' textbook type length scales
do not increase as dramatically as the relaxation time does when a liquid is supercooled and becomes a glass.  

It is worthwhile to highlight that one of the most pertinent naturally increasing length scales is that associated with the typical size of the dynamical
heterogeneity (i.e., $\xi_{4}$ of Eq. (\ref{Scl})). Similar to other measures, this typical length scale does not increase as rapidly as the characteristic
relaxation time does as the glass transition is approached. Recent work for a three-dimensional bi-disperse repulsive glass \cite{mizuno}
of Eq. (\ref{bi-disperse}) 
suggests that 
\begin{eqnarray}
\label{tauxi}
\tau \sim \exp(k \xi_{4}^{\theta}),
\end{eqnarray}
with $\theta \simeq 1.3$ and $k$ a constant. 
An alternate assumed algebraic form $\tau \sim \xi_{4}^{z}$ leads to a large dynamical exponent $z \simeq 10.8$.
In these cases, the dynamics changes dramatically with little notable change in the spatial correlation length.

\subsection{Quantum Critical Jamming}
\label{sec:jamming}

The mapping between dissipative classical and quantum systems raises the specter of a new
quantum critical point associated with jamming \cite{physics_ZN}. Our discussion below employs the exact mapping of Eq. (\ref{qcqc}) to derive exact results on quantum jammed systems given the wealth of 
information on their classical counterparts. 

The classical jamming transition \cite{J,arcm,J1,J2,J3,silbert,J4,J5,J6,J7,J8,hatano} of hard spheres
(such as those of models (A) in section \ref{models})
from a jammed system at high density to an unjammed one with spatially heterogeneous motion at lower densities
is a continuous transition with known critical exponents, both static
\cite{J3,silbert} and dynamic \cite{hatano}. The transition into the classical jammed phase may be 
brought about by a reduction in temperature, increase in particle density and the application of stress. 
In most classical solids, the
ratio of the shear to bulk modulus is a number of order unity (e.g., $1/3$ in many conventional three-dimensional solids). However, at
the jamming threshold, this ratio tends to zero. 
Thus, jammed system may be very susceptible to shear stresses. This softness is one of the peculiar features that sets jammed systems apart from conventional
solids \cite{arcm}. As seen by our mapping from classical to quantum systems, the classical jamming transition has a quantum analog with similar dynamics.
Replicating the mapping of the previous subsection (and, in particular, Eq. (\ref{ggqc}) therein),
we may derive an analog quantum 
system harboring a zero temperature transition
with similar critical exponents. The classical zero temperature
 critical point (``point  J")  \cite{J, J1} 
may rear its head anew in the form of {\em quantum critical jamming} (at a new critical point $-$ ``quantum point J'')
in bosonic systems. A schematic of the phase diagram of the associated quantum system
is depicted in Fig. \ref{J-point_fig}.

 We may ascertain dynamical exponents from
those reported for the classical jamming system \cite{hatano}. The classical low temperature temperature system ($T_{cl} \to 0$) maps, according to Eq. (\ref{HQM}), onto a zero temperature 
quantum system in its large mass limit. 
Bosons of infinite mass are not trivial due their statistics. Specifically, for a classical system of mono-disperse soft spheres
with a repulsive force that is linear in the amount of compression, it was found that the correlation length $\xi$ and relaxation time $\tau$
scale \cite{hatano} as 
\begin{eqnarray}
\xi \sim (\rho_{J} - \rho)^{-0.7}, \nonumber
\\ \tau \sim (\rho_{J} -  \rho)^{-3.3}.
\label{jamm}
\end{eqnarray}
In Eq. (\ref{jamm}), $\rho$ denotes the density with $\rho_{J}$ being the critical density at the jamming transition marked by point $J$.
Eq. (\ref{jamm}) describes how the spatial and time scales diverge as the density is increased and approaches (from below)
the density at the jamming transition. The correlation length in the jammed systems is set by the scale at which the number of surface zero modes is balanced by bulk effects.
Taken together, these imply that, on approaching the transition, the relaxation time increases much 
more rapidly than the correlation length, $\tau \sim \xi_{4}^{z}$ with a large effective dynamical exponent $z \simeq 4.6$.  
By use of Eq. (\ref{ggqc}), the same behavior is to be expected for the quantum system 
governed by the corresponding quantum potential ${\cal{V}}_{\sf Quantum}$. 
The classical (and thus quantum) jamming exponents are the same in two and three-dimensions.

It may be remarked that a similar dynamical exponent was found for a Bose glass model suggested to describe vortex lines in high temperature superconductors \cite{another4.6}. 
In physical terms, for charged bosons, the jamming transition constitutes 
a transition from a metallic system (when the system is unjammed
and behaves as a fluid) to a jammed state (an insulator). 
We note earlier work rationalizing metal to insulator
transitions in terms of electron pinning \cite{reichhardt2004}.
In the bosonic jamming that we describe above,
no pinning is present and the transition is
driven by particle interactions.

For completeness, we make one further remark concerning the physics of the jamming transition and its relation
to the glass transitions that we discussed hitherto. As found in \cite{ludovic,ludovic1}, the density $\rho_J$ is an important density as it corresponds to a change in the properties of the glass phase. The conventional jamming
transition does not correspond to a transition into a glass. Rather, point $J$ and its finite temperature extension lies deep within a glassy phase
that already onsets at a far lower density. The jamming transition at point $J$ is associated with changes in the mechanical/structural properties of the glass phase. 
It is this transition that we depict in Fig. \ref{J-point_fig}.

As in earlier sections, we see that time scales increase far more
precipitously than spatial correlation lengths. 

\begin{figure}[h]
\includegraphics[angle=0,width=1.0\columnwidth]{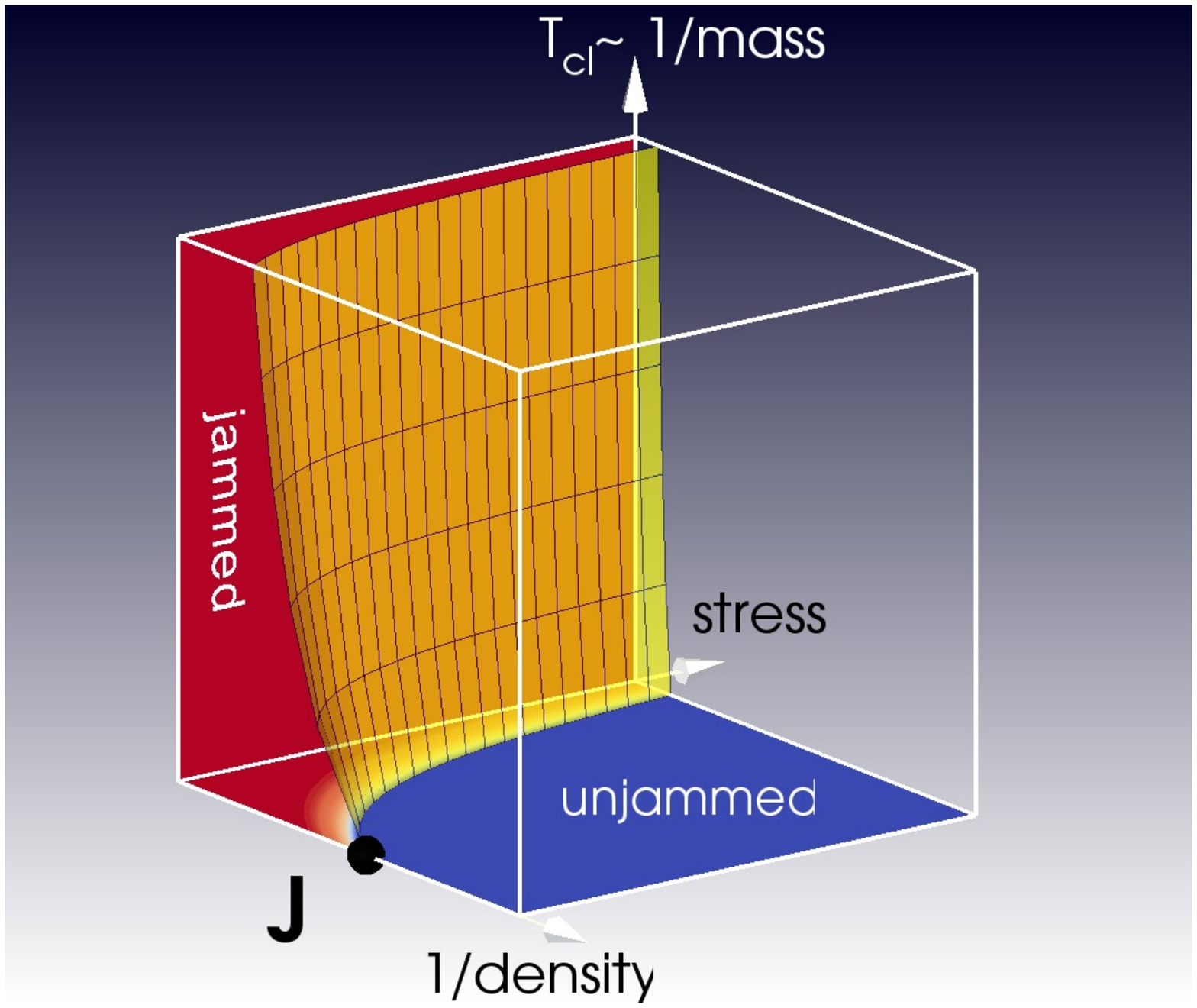}\\
\caption{(Color online.)
The phase diagram of the zero temperature quantum jamming transition with line of $J$ points.
The phase boundaries and axis were formed by employing the
phase diagram of the classical system \cite{arcm} and examining the duality between 
the classical and quantum system- i.e., comparing the parameters
in the quantum system of  Eq. (\ref{HQM}) with the classical system defined by 
Eqs. (\ref{Langevin},\ref{noiseeq}).
}
\label{J-point_fig}
\end{figure}

One of the hallmarks of classical jammed systems is that the spectral density of vibrational excitations, $D(\omega)$ is constant at the
jamming threshold \cite{silbert,zer}.  (In conventional Debye solids, $D(\omega) \sim \omega^2$.)  This near constant value of $D(\omega)$ in jammed systems is
independent of potential, dimension, and size of the system.  Away from the jamming threshold, $D(\omega)$ exhibits a plateau down to a frequency
$\omega^*$.  Below $\omega^*$, the system behaves similar to a Debye
solid.  This crossover frequency $\omega^*$ veers to zero on approaching the
transition.  The number of low energy modes is set by the absence of constraints
and Maxwell counting arguments. This constant density of states implies an enormous increase in the
low frequency excitations.  These excitations can be probed by examining the trajectory of a
single particle and Fourier transforming its motion.  The corresponding modes
are quasi-localized (or resonant) at low frequency below $\omega^*$.  Above
$\omega^*$, these modes are extended but still do not look at all like plane waves.  \cite{arcm}

We now turn to the quantum systems derived by the mapping of Eq. (\ref{HQM}). 
The results concerning the density of the modes in classical systems hold unchanged for their dual quantum
counterparts. This is so as the dynamical matrix $D$ is formed by the second derivatives of the potential
$V_{\sf N}$ relative to the displacements. Thus, the same statement about mode density of
states that appear in classical jammed systems holds verbatim in the classical potential
$V_{\sf N}$ which we use to construct the quantum many body potential  ${\cal{V}}_{\sf Quantum}(\{\vec{x}\})$ 
from Eq. (\ref{HQM}). It follows that any appearance of zero energy (bulk or surface) modes in the classical system
will identically hold also in the quantum system.

\section{Bosonic Lattice systems}
\label{ls}

Thus far, we largely focused on continuum viscous classical systems which, as we have seen, mapped onto continuum bosonic systems. We briefly remark here
on classical lattice systems which similarly exhibit dynamical heterogeneities and a jamming type transition. 
[In subsection \ref{latfi}, we further
briefly expanded on the extension of our derivation for lattice systems.] Refs.\cite{2DN3,3DN2} studied, respectively,
the 2DN3 and 3DN2 models on the square and cubic lattice models in $d=2$ and $d=3$ dimensions. 
We provided details for these lattice models in the discussion of models of class (D) in section \ref{models}.
Similar to the continuum systems that we largely focused on until now, these models may be regarded as those of classical hard core spheres.
These finite range hard core interactions on the lattice thwart crystallization
and lead to an amorphous jammed phase at high density.
Following the mapping reviewed in section \ref{rev-sec}, the quantum Bose counterpart of such systems is that of dominant hard sphere interactions augmented by contact sticky interactions.
In the classical systems, simulation starts \cite{2DN3,3DN2} with an infinitely fast quenching wherein particles are added whenever possible and diffuse otherwise;
this process is halted when the desired density is reached.  A clear increase was noted in the length scales that characterize the dynamical heterogeneity \cite{2DN3,3DN2}. 
The continuum jamming transition discussed earlier may have a lattice counterpart for Cooper pairs as we now elaborate on. The jammed phase is that an insulating
(or Mott) phase of hard core bosons forming a Hubbard type system. 
Specifically, a natural quantum counterpart to the N3 (N2) model is given by an extended Bose Hubbard \cite{mpa,zim} type model with infinite hard core repulsions,
\begin{eqnarray}
\label{HNM}
H = - t \sum_{\langle ij \rangle}  (b_{i}^{\dagger} b_{j} + h.c.) + U \sum_{i} n_{i}(n_{i}-1)  \nonumber \\
 + \sum_{ij} V_{ij} n_{i} n_{j},
\end{eqnarray}
where $V_{ij} \to \infty$ for lattice sites $i$ and $j$ which are fewer than four (or three) steps apart and the onsite 
Hubbard repulsion $U$ is divergent ($U \to \infty$) as well. The Hubbard term leads to a penalty only when 
there is a double or higher occupancy. Based on our considerations thus far, we expect to
obtain the quantum bosonic counterpart to the classical jamming transitions found in the classical 
2DN3 and 3DN2 models. This bosonic system may have all of the characteristics of the 
classical jammed system including dynamical heterogeneities and a large
dynamical exponent $z$. For completeness, we briefly comment on the
difference between the lattice system of Eq. (\ref{HNM}) and 
the ``Bose glass''  first introduced in \cite{mpa}. The Bose glass appears in
the bare (i.e., that with $V_{ij}=0$) disordered rendition of Eq. (\ref{HNM})
with the general Bose Hubbard Hamiltonian (with general finite repulsion $U$) 
being further augmented by a local chemical potential term $-  \sum_{i} \mu_{i} n_{i}$ wherein 
$\mu_{i}$ is a spatially non-uniform random quantity. By contrast, the lattice Hamiltonian of Eq. (\ref{HNM}) as 
well as the continuum models that we discussed in earlier sections are free of disorder. The amorphous characteristics 
that these clean systems may exhibit are borne out of ``self-generated'' randomness \cite{sw}---not randomness that is present in the parameters
defining the system.

\section{Electronic systems with pairing interactions}
\label{esp}

Up to now, building on and extending the mapping between classical dissipative systems and zero temperature bosonic theories, 
we focused on hard core bosons. We now turn to the ground states of Fermi systems. 
In particular, in this section, we will consider standard electronic systems with pairing interactions,
\begin{eqnarray}
H = \sum_{\vec{k}, \sigma} \epsilon_{\vec{k}} c_{\vec{k} \sigma}^{\dagger} c_{\vec{k} \sigma} + 
\sum_{\vec{k}, \vec{l}} V_{\vec{k},\vec{l}} c_{\vec{k} \uparrow}^{\dagger} c_{-\vec{k} \downarrow}^{\dagger} c_{-\vec{l} \downarrow} c_{\vec{l} \uparrow},
\label{BCS+}
\end{eqnarray}
where $\sigma = \uparrow, \downarrow$ is the spin polarization index and $V_{kl}$ is the interaction strength between the Cooper pairs \newline
 $|\vec{k} \uparrow; -\vec{k} \downarrow \rangle$ and $|\vec{l} \downarrow ; - \vec{l} \uparrow \rangle$. 
As is well known (and is readily verified), the following Fermi billinears 
\begin{eqnarray}
\label{bbb}
\overline{b}_{\vec{k}}^{\dagger} = c^{\dagger}_{\vec{k} \uparrow} c^{\dagger}_{-\vec{k} \downarrow}, \nonumber
\\ \overline{b}_{\vec{k}} = c_{-\vec{k} \downarrow} c_{\vec{k} \uparrow},
\end{eqnarray}
corresponding to the creation/annihilation of Cooper pairs satisfy hard core Bose algebra. We next consider what occurs if, within the ground state, 
the occupancies of the single particle states are correlated inasmuch as the electronic states
on which the standard pairing Hamiltonian of Eq. (\ref{bbb}) operates
can be created by applications of Cooper pair creation operators on the vacuum (i.e., if
the ground state is invariant under the combined operations of parity ($ \vec{k} \to -\vec{k}$) 
and time reversal ($\sigma \to -\sigma$)). When the ground state is strictly invariant under the combined
effect of these symmetries, we may express the Hamiltonian of Eq. (\ref{BCS+}) as a bilinear in the 
hard core Bose operators, 
\begin{eqnarray}
\label{XYM}
H =  \sum_{\vec{k},\vec{l}} (2 \epsilon_{\vec{k}} \delta_{\vec{k},\vec{l}} + V_{\vec{k},\vec{l}}) \overline{b}_{k}^{\dagger} \overline{b}_{l}. % \equiv \sum_{\vec{k} \vec{l}} h_{\vec{k} \vec{l}} \overline{b}_{\vec{k}}^{\dagger} \overline{b}_{\vec{l}} .
\end{eqnarray}
The hard core (Fourier space) Bose algebra of the creation and annihilation operators [as, in particular, manifest in
the relation $(\overline{b}_{\vec{k}}^{\dagger})^{2} =0$ mandating that no more than one boson can occupy any given (Fourier space) site]
is identical
to that of raising and lowering operators in the spin $S=1/2$ system. Thus, a simple extension of the standard real-space Matsubara-Matsuda transformation \cite{MM} 
is given by
\begin{eqnarray}
\label{matsuda}
\overline{b}_{\vec{k}}^{\dagger} \to S_{\vec{k}}^{+}, \nonumber
\\ \overline{b}_{\vec{k}} \to S_{\vec{k}}^{-}.
\end{eqnarray}
Substituting Eq. (\ref{matsuda}) into Eq. (\ref{XYM}), we arrive at an XY model.
In situations in which the band dispersion $\epsilon_{\vec{k}}$ is nearly flat (and may be omitted for fixed particle number), in determining the ground state(s),
we must only find the pairing $V$ that affects pair hopping.  Similar considerations apply in real space when Cooper pairs are short ranged and may be replaced by real-space hard-core bosons. 
Hard core real space contact interactions correspond to uniform $V_{\vec{k},\vec{l}}$ (independent of $\vec{k}$ and $\vec{l}$) as in the BCS form for the pairing interactions.   
In such cases, whenever the system is dominated by hard core contact interactions between the bosonic Cooper pairs we see, replicating our analysis thus far, at zero temperature, 
that the system may undergo a jamming type transition between an itinerant and jammed phase at sufficiently high densities or pressure. 
In this case, it displays rapidly increasing relaxation times concomitant with spatial correlations that do not increase as dramatically
as the relaxation times do on approaching this transition.

\section{Possible implications for experimental data}
\label{exp_implications}
We now, very briefly, turn to a discussion of possible data analysis of experiments.
One of the main messages of our work is that classical physics associated with 
overdamped classical systems can rear its head in the quantum arena.
Correspondingly, data analysis which has led to much insight
in the study of classical glasses and other damped 
systems may be performed anew for quantum
systems. A principal correlation function
which we focused on in this work 
has been that of the four-point correlation
function of Eq. (\ref{g4xy}). This correlation function
need not be directly measured in real time. 
For instance, scanning tunneling spectroscopy (STS)
data taken at different positions and bias voltages may provide
a valuable conduit towards the evaluation
of the four-point correlator
when it is expressed as
an integral over frequencies 
(or associated bias voltages).
Rather trivially with $\phi(\vec{x},V)$ denoting the local density
of states at location $x$ for a bias voltage $V$, and $e^*$ the 
electronic charge,
the corresponding four-point correlation function is given by
\begin{eqnarray}
G_{4}(\vec{x}-\vec{y},t) = \int dV_{1} dV_{2} dV_{3} dV_{4} \nonumber
\\  \langle \delta \phi(\vec{x},V_{1})\delta \phi(\vec{x},V_{2}) \delta \phi(\vec{y},V_{3}) \delta \phi(\vec{y},V_{4})\rangle 
e^{i e^{*} t (V_{1}+V_{3})} \nonumber
\\ - C(\vec{x},t) C(\vec{y},t) ,
\end{eqnarray}
with the two-point auto-correlation function
\begin{eqnarray}
C(x,t) = \int dV
\langle \delta \phi(\vec{x},V)\delta \phi(\vec{x},0)\rangle e^{i e^{*} t V}.
\end{eqnarray}
As seen from our discussion in Section \ref{qdhsub}
concerning the Fourier transformed four-point correlation function $S_4^{classical}(\vec q, t)$ in Eq. (\ref{Scl}),
quantum dynamical heterogeneities may be manifest in this
correlation function.

\section{Conclusions}
\label{conc}
A central result of this work is the exact temporal correspondence of Eq. (\ref{ggqc}) that spans both equilibrium and non-equilibrium dynamics
in general time dependent systems
(so long as either the initial or the final state of the classical system is that of thermal equilibrium).  This equality relates {\bf (i)} the auto-correlation function of Eq. (\ref{gclq}), for {\em any quantity} ${\cal{O}}$ 
when evaluated for the classical dissipative system of Eq. (\ref{Langevin}) with a many body potential energy $V_{N}$, 
 to {\bf (ii)} the auto-correlation function of the very same corresponding quantum operator $\hat{\cal{O}}$
in a dual bosonic system governed by the Hamiltonian of Eq. (\ref{HQM}). When fused with known results for dissipative classical 
systems, this extremely general equality immediately leads to numerous non-trivial effects which we introduced and readily proved as a matter of principle. These include:

$\bullet$  {\em Quantum dynamical heterogeneities} (QDH). We illustrated that similar to classical systems even in the absence of disorder, 
bosonic systems can, at zero temperature, exhibit spatially non-uniform zero-point motion. Of course, in translationally invariant systems, the average (time averaged) dynamics is uniform.
However, at any given time, there are particles that move more rapidly than others. We suggested how experimental data may be analyzed to search 
for quantum dynamical heterogeneities in electronic systems.

$\bullet$ The length scale characterizing the zero temperature QDH, the four-point correlation length $\xi_{4}$ (a trivial analog of its classical counterpart) may increase as the dynamics of the clean Bose system becomes progressively sluggish. 
However, albeit its rise, this length scale may increase much more slowly than the relaxation time. The far more rapid increase of the relaxation time as compared to readily measured length scales is 
a hallmark of many electronic systems. Cast in terms of quantum critical scaling (if and when it might be realized), the effective dynamical exponent $z$ capturing the relation between correlation lengths and times
is very large ($z \gg 1$). Other relations such as those of Eq. (\ref{tauxi}) may hold once they are established for viscous classical systems. I

$\bullet$ The dramatic increase of primary relaxation times 
(which are far larger than the increase in conventional static length scales) with classical temperature as given by Eq. (\ref{VFT_eq}) [as well as the secondary relaxations of Eq. (\ref{secondary})]
have direct zero temperature quantum analogs wherein changes in the classical temperature are replaced by a scaling of effective mass of particles and form of the man body potential ${\cal{V}}_{\sf Quantum}(\{\vec{x}\})$.

$\bullet$ Classical stretched exponential relaxations of the form of Eq. (\ref{stretch}) have quantum analogs in the form of Eq. (\ref{qqq}). In the quantum arena, there are sinusoidal modulations that multiply 
stretched exponential type relaxations.

$\bullet$ Similar to classical systems, quantum systems may jam at high densities or pressure notwithstanding zero point motion. The character of the jamming transition in zero temperature quantum systems
is identical to that of their corresponding classical finite temperature counterparts. As the classical systems exhibit a critical point at the jamming transition (at ``point J'')  so do their bosonic counterparts. 
As a result, we established the existence of a new quantum critical point---associated with a {\em quantum critical jamming} of a hard core Bose system. As in the other systems that we discussed, the characteristic relaxation time 
diverges more precipitously than the correlation length on approaching the transition (``quantum point J'') with a large effective dynamical exponent $z \simeq 4.6$. 

$\bullet$ The continuum theories that we predominantly focused on may have a broad applicability as continuum theories describe the same physics as their lattice renditions do in the vicinity of critical points. 
In Section \ref{ls}, we discussed specific possible lattice renditions.

$\bullet$ The results that we derived for zero temperature bosonic theories suggest similar features in electronic systems. In some cases, as discussed in Section \ref{esp}, finding the ground states of interacting electronic systems can be cast in terms of a corresponding zero temperature hard-core Bose problem.
 
 $\bullet$ The general mapping of Eq. (\ref{matsuda}) between hard core bosons to $S=1/2$ spin systems (in either momentum ($\vec{k}$) or real ($\vec{r}$) space) along with 
 the complementary relation for the $z$ component of the spin, 
 \begin{eqnarray}
 [\overline{b}_{\vec{k}}^{\dagger} \overline{b}_{\vec{k}} -1/2] \to S^{z}_{\vec{k}},
 \end{eqnarray}
 in the same space, allows us to derive similar results for certain spin $S=1/2$ systems. Spin models
 may exhibit transitions from spin-liquid type phases to disorder free glassy systems. In
 these systems, dynamical heterogeneities concomitant with a notable increase in relaxation time scales may arise.

Thus, with the aid of the viscous classical many body quantum correspondence of
Eq. (\ref{ggqc}), we trivially established all of these results without the need to perform various standard and far more laborious computations for quantum systems.

Other possible extensions of our results include the relation between
localization (or caging) in classical systems and their corresponding
quantum counterparts. We may similarly examine disordered systems $-$ 
for a random classical potential $V_{\sf N}$, the corresponding 
quantum potential ${\cal{V}}_{\sf Quantum}$ is also random. 
Although our focus has been on supercooled systems, Eq. (\ref{ggqc}) implies that also standard (non glassy or spin-glassy) classical transitions
have corresponding zero temperature quantum analogs. 

As we discussed in detail (see Eq. (\ref{psie})),
 if we are given a known quantum ground state,
 we may find the corresponding effective classical potential. With the aid of calculations on how the correlation functions
 of the classical system depend on time as parameters in the classical potential are varied, we may then 
 determine the corresponding time dependent correlation functions of the dual many body quantum system. 
 That is, we need not always find corresponding quantum systems to classical 
 systems; the Fokker-Planck mapping also enables to go in the opposite direction
 from quantum systems to classical ones. 
 
Numerous related extensions may be considered. For instance, we may consider magnetic and other systems in which fermionic degrees can be formally integrated out leaving
only effective bosonic degrees of freedom.  Consequences for the Ward identity relating four-point with two-point correlation functions (as in, e.g., \cite{BK})
may be considered. The Langevin equation may be re-examined for single vortex crossing of a narrow superconducting wire at finite temperature 
to derive the mapping for the quantum dual at absolute zero temperature.\cite{Bulaevskii2011,Bulaevskii2012} 
This would offer an alternative path for exploring the viability of quantum phase slips in nanowires.

%{\bf Acknowledgements.} 
\acknowledgments

ZN thanks Mark Alford, Boris Altshuler, 
Carl Bender, John Cardy, Seamus Davis, Silvio Franz, Victor Gurarie, Alioscia Hamma, Eun-Ah Kim, Andrea Liu, Sid Nagel, and Charles Reichhardt for discussions and ongoing work.
In particular, section \ref{exp_implications} was triggered by a question raised by Eun-Ah Kim and Seamus Davis. Work at Washington University in St. Louis has been supported by the National
Science Foundation (NSF) under Grant number NSF DMR- 1106293. Research at the KITP was supported, in part, by the NSF under Grant No. NSF PHY11-25915. ZN also thanks Los Alamos National Laboratory (LANL) where a  part of this work was done. ZN and AVB further thank the Aspen Center for Physics for hospitality and NSF Grant No. 1066293.
Work at LANL was carried out under the auspices of the 
NNSA of the U.S. DOE at LANL under Contract No.\
DE-AC52-06NA25396 through the Office of Basic Energy Sciences, Division of Materials Science and Engineering.

\appendix

\section{Analytic continuation of classical stretched exponentials}
\label{dets}

 In subsection \ref{TwoOperators}, we explicitly illustrated how a Wick type rotation 
 $t \to i t$ relates time dependent correlation functions in the viscous classical systems to 
 those in the dual quantum many body theories. 
 For the sake of clarity, we explicitly discuss how the analytic continuation of the classical correlation function should be performed
 in some simple yet empirically relevant cases for dynamical response functions
 in viscous classical systems wherein Eq. (\ref{stretch}) describes the dynamical response. 
In Eq. (\ref{qqq}), we provided the quantum dual to exponentially stretched 
classical dynamics. In this brief appendix, we describe how this result
is derived and outline how analytic continuations for other 
response functions of classical viscous systems may be
analytically continued following such a Wick type rotation.  
We thank Carl Bender for a quick tutorial on aspects of Stokes' wedges
on which this brief appendix heavily relies.

If $G_{classical}(t) = \sum_{n} B_{n} e^{-t/\tau_{n}}$ then, trivially,
the quantum response function will be uniquely defined and given by
$R_{Q} = \sum_{n} B_{n} \cos (t/\tau_{n})$. The same applies, of course,
for distributions of modes (whence the discrete sum over overdamped classical
modes $n$ is replaced by an integral with some density of modes $f(\tau)$). 
In the limit of an infinite number of modes,

\begin{eqnarray}
G_{classical}(t) = \int_{0}^{\infty} d \tau' f(\tau') \exp(-t/\tau'),
\label{lnggcl}
\end{eqnarray}
with $f$ a distribution that generally is no longer a sum of Dirac delta functions. 
As stated in Eq. (\ref{qcqc}), 
$G_{quantum}(t) = G_{classical}(it)$. Thus, 
\begin{eqnarray}
G_{Quantum}(t) =  \int_{0}^{\infty} d \tau' f(\tau') \exp(-it/\tau').
\label{lngqu}
\end{eqnarray}
In the complex $\tau'$ plane, for any ``well-behaved'' function $f(\tau')$ that is localized in a region of positive finite $\tau'$, the integral of Eq. (\ref{lnggcl}) may be performed
along any contour connecting the origin and the $\tau' =\infty$ along the real line such that the contour lies exclusively in the right half complex plane (the pertinent Stokes wedge in this case) of a positive real component of $\tau'$, i.e., $\Re \{ \tau'\} \ge 0$. 
Of particular interest to us is the stretched exponential form given by $G_{\sf classical}(t) =A \exp[- (t/ \tau)^{c}]$.
Now, in performing the substitution $t \to it$ to implement the transformation from Eq. (\ref{lnggcl}) to Eq. (\ref{lngqu}), we perform the rotation 
\begin{eqnarray}
\label{rot}
t = t e^{i \varphi}, ~ ~\varphi:~ 0 \longrightarrow \varphi_{final},
\end{eqnarray} with 
$\varphi_{final} = \pi(4n+1)/2$ where $n$ is an integer.
The integral of Eq. (\ref{lngqu}) remains well-defined in the top complex half-plane of a positive real part of $\tau'$, i.e.,  $\Re \{\tau'\}  \ge 0$ (the rotated counterpart of the original Stokes wedge). If $\varphi$ is varied continuously from $0$ to $\pi/2$, there remain contours from $\tau'=0$ to $\tau'= \infty$
that appear in the original Stokes wedge of $\Re \{ \tau'\} \ge 0$
that pass exclusively through the region $ \Im \{\tau'\}  \ge 0$; the integrals along these contours can be analytically continued 
when $\varphi$ is continuously increased from $0$ to $\pi/2$. Thus, we may perform the rotation of Eq. (\ref{rot}) continuously increasing $\varphi$
to represent $i$ as $e^{i \pi/2}$ and replace $t \to t e^{i \pi/2}$ in the argument of $G_{classical}(t)$. This is what we have done in Eq. (\ref{qqq}).
For other choices of $n$ for $\varphi_{final} = \pi(4n+1)/2$ as we continuously vary $\varphi$ from its initial value of zero, 
there will always appear situations where the original Stokes wedge will have no overlap with its rotated counterpart. Thus, the substitution of $t \to t e^{i \pi/2}$ in the argument of $G_{classical}(t)$
is the only one that may be implemented out of the possible choices in Eq. (\ref{rot}) in order to evaluate the integral of Eq. (\ref{lngqu}). 
This forms the correct analytic continuation of the original real time correlation function of $G_{classical}(t)$ of Eq. (\ref{lnggcl}).

\section{Relation between the classical and quantum potentials in the eikonal approximation to the Schr\"{o}dinger equation}
\label{sec:action}

Below we briefly review the eikonal approximation and then discuss its relation to the connection between the classical and quantum many body
potentials as seen in Eqs. (\ref{HQM}, \ref{psie}). This link lies at the heart of Madelung Hydrodynamics \cite{madelung}.Towards this end, we write the wavefunction as a function of only the phase, the eikonal approximation [as throughout, we set $\hbar=1$],
\begin{eqnarray}
\Psi_{0}= A e^{i S},
\label{eikonal}
\end{eqnarray} 
and substitute this into the Schr\"{o}dinger equation with the Hamiltonian in the second line of Eq. (\ref{HQM}) then we will arrive at
\begin{eqnarray}
\frac{1}{2m} \sum_{i} (\vec{\nabla}_{i} S)^{2} +  {\cal{V}}_{\sf Quantum}(\{\vec{x}\}) + \frac{\partial S}{\partial t}\nonumber
\\ =  \frac{i}{2m}\sum_{i} \nabla_{i}^{2} S.
\label{eikonal_eq}
\end{eqnarray}
For time independent solutions, $\frac{\partial S}{\partial t}=0$ and Eq. (\ref{eikonal_eq}) rather trivially becomes
\begin{eqnarray}
 {\cal{V}}_{\sf Quantum}(\{\vec{x}\}) = \sum_{i} [ \frac{i}{2m}\nabla_{i}^{2} S- \frac{1}{2m} (\vec{\nabla}_{i} S)^{2}].
 \label{VQS}
 \end{eqnarray}
 If we now invoke the correspondence $iS \leftrightarrow -\beta  V_{\sf N}/2$ then 
 Eq. (\ref{eikonal}) will transform into Eq. (\ref{psie}) and, similarly, Eq. (\ref{VQS}) will become
 Eq. (\ref{HQM}) relating the quantum potential energy ${\cal{V}}_{\sf Quantum}$ to
 the classical potential energy $V_{\sf N}$.

\section{Simple examples of classical to quantum correspondence and their aspects}
\label{ssimple}

To elucidate some aspects of the known mapping between classical dissipative and quantum 
systems reviewed in subsection \ref{FPreview}, we discuss several extremely simple examples in $d$ spatial dimensions. 

\subsection{Non-interacting particles}
\label{fps}

For a (free) system having zero potential everywhere, the quantum ground state wave-function is a constant in real space. 
That this is so can be seen by our mapping and the form of the classical probability in Eq. (\ref{psie}) for vanishing classical potential energy.
By Eq. (\ref{HQM}) the same also occurs for the quantum potential, which is everywhere zero: ${\cal{V}}_{\sf Quantum} =V_{\sf N} =0$.

\subsection{Zero energy bound state}
For a short range attractive potential
the zero energy eigenstate outside the potential, up to volume normalization factors, 
given by
\begin{eqnarray}
\Psi_{0}(\vec{x}) = \frac{A}{|\vec{x}|^{d-2}}. 
\end{eqnarray}
Invoking Eq. (\ref{psie}), we see that, in this case, 
\begin{eqnarray}
\label{vnf}
V^{\sf free}_{\sf N}(\{\vec{x}\}) =2T_{cl}(d-2) \ln |\vec{x}|.
\end{eqnarray} 
Indeed substituting Eq. (\ref{vnf}) into Eq. (\ref{HQM}) and recalling that, in its scalar ``S-wave'' (or ``$\ell =0 $'') representation, the Laplacian is 
given by $\nabla^{2} = \frac{d^{2}}{dr^{2}} + \frac{d-1}{r} \frac{d}{dr}$, it is readily verified, as it must self-consistently be, that the corresponding quantum potential ${\cal{V}}_{\sf Quantum} =0$
in the region outside the range of the interaction.

\subsection{Harmonic oscillator systems}
\label{hs}
As seen by Eq. (\ref{HQM}), classical systems with harmonic potentials  $V_{\sf N}$ map onto quantum systems with similar (up to innocuous shifts) harmonic potentials ${\cal{V}}_{\sf Quantum}=V_{\sf N} + {const}$.
That this must be so is readily seen as the ground state $\Psi_{0}$ of simple quantum harmonic potentials is given by a Gaussian.  Using 
Eq. (\ref{psie}), we see that this indeed relates to a harmonic classical potential $V_{\sf N}$ as it must. As can be further seen from Eq. (\ref{saa}), in the case
of harmonic classical systems, the operators $A$ and $A^{\dagger}$ are trivially related to the raising and lowering operators
in the quantum harmonic problem (and indeed the Gaussian form of the ground state can, as is very well known, be seen 
from the requirement that the annihilation operator must yield zero when acting on the ground state).

\subsection{Scaling invariance of time and space}

As is well known, for a homogeneous classical potential $V_{\sf N}(\{\vec{x}\})$ which scales as a power (say, $p$) of the spatial coordinates $|\vec{x}|$,
the equations of motion are invariant under a simultaneous rescaling of the time coordinates. This analysis is typically done for inertial systems. 
When replicated for the over damped system of Eq. (\ref{Langevin}), we find that 
\begin{eqnarray}
\label{scale_inv}
\vec{x}_{i} \to a \vec{x}_{i}, ~ t \to b t,
\end{eqnarray}
where $b$ plays the role of $\lambda$ and $a$ plays the role of $\lambda^{1/z}$ from before,
leads to an invariance of Eq. (\ref{Langevin})
if $b = a^{2-p}$. 
By contrast, in the corresponding quantum problem of the Schr\"{o}dinger equation with the Hamiltonian of Eq. (\ref{HQM}), a scaling such as that of 
Eq.(\ref{scale_inv}) is possible only for a single case: that of a potential $V_{\sf N}(\{\vec{x}\})$ that 
is a logarithmic function of its arguments (or a constant). For this particular case, we find
that $b=a^{2}$. Correspondingly, akin to subsection \ref{fps}, 
for this particular case, the time scales as $t \sim |x|^{2}$ as in diffusion
or the free particle quantum problem.

\section{Slater-Jastrow forms}
\label{sslater}

The general results presented thus far may, in some instances, be generalized to describe fermions. A limited 
extension is the one concerning the evolution starting off from an initial Slater-Jastrow type fermionic wavefunction. As we have emphasized
earlier, if the $V_{ij}$ in Eq. (\ref{vnc}) are symmetric under the exchange of $i$ and $j$, the resulting wavefunction obeys Bose statistics.  This symmetry is 
maintained for the ground state as it is a Jastrow function given by Eq. (\ref{psie}).  Fermionic wavefunctions are afforded by the product 
of the symmetric boson ground state and an antisymmetric term,
\begin{eqnarray}
\Psi_{F} = \Psi_{0} \chi.
\label{GeneralFermionWavefunction}
\end{eqnarray}

The function $\chi$ can take any antisymmetric form. For simplicity, we choose it to be
a Slater determinant of the form 

\begin{eqnarray}
\chi = \frac{1}{ \Omega ^{N/2}} \frac{1}{\sqrt{N!}} \left|
\begin{matrix}
    e^{i \vec{k}_1 \cdot \vec{r}_1} & e^{i \vec{k}_1 \cdot \vec{r}_2} & \cdots &e^{i \vec{k}_1 \cdot \vec{r}_N} \\
   e^{i \vec{k}_2 \cdot \vec{r}_1} &e^{i \vec{k}_2 \cdot \vec{r}_2} & \cdots &e^{i \vec{k}_2 \cdot \vec{r}_N} \\
    \vdots               & \vdots               &        & \vdots               \\
   e^{i \vec{k}_N \cdot \vec{r}_1} & e^{i \vec{k}_N \cdot \vec{r}_2} & \cdots & e^{i \vec{k}_N \cdot \vec{r}_N}
\end{matrix} \right|,
\label{SlaterDeterminant}
\end{eqnarray}
with $\Omega$ the volume of the system.

As $i \partial_t \Psi_0 = H \Psi_0$,

\begin{eqnarray}
i \partial_t (\Psi_{0} \chi) &=& \chi (i \partial_t \Psi_0) + \Psi_0 (i \partial_t \chi)
\nonumber\\ 
%\;\;\;\;\; = \chi H \Psi_0 + \Psi_0 E_{\sf Slater} \chi
%\nonumber\\ 
 &=& \chi \left[ H + \ E_{\sf Slater} \right] \Psi_0,
\end{eqnarray}
where $E_{\sf Slater}$ is the energy of the free particle system described by $\chi$ and
\begin{eqnarray}
H ( \Psi_{0} \chi) = (T_0 + {\cal{V}}_{\sf Quantum})
(\Psi_{0} \chi).
\end{eqnarray}
The potential energy operator, in the second term, leads to $\chi ( {\cal{V}}_{\sf Quantum} \Psi_0)$.  The kinetic energy operator $T_0$ generates three terms of, respectively, the forms 
$\sum_{a}(\nabla_a^2 \chi) \psi^0$, $\sum_a (\nabla_a \chi) \cdot (\nabla_a \psi^0)$, and $\sum_a \chi (\nabla_a^2 \psi)$.  The first and the 
last of these terms represent the term proportional to $E_{\sf Slater}$ and the original bosonic kinetic energy respectively.
The second term, that of the mixed gradients, is proportional to $\sum_a \vec{k}_a$.
For a system invariant under parity, this sum vanishes. Up to an innocuous phase factor, the evolution given an initial fermionic wavefunction of Eq. (\ref{GeneralFermionWavefunction}), will be thus identical to that with the bosonic wavefunction $\Psi_{0}$ and all correlation functions will be identical to those which we earlier computed for the bosonic system. That is, the general time dependent correlation functions given an initial fermionic
state of Eqs.(\ref{psie},\ref{GeneralFermionWavefunction}) will adhere to the general $t \to it$ rule which we detailed in earlier sections. A notable difference by comparison to the bosonic case, however, 
is that the wavefunction of Eq. (\ref{GeneralFermionWavefunction}) at an initial (or at a final) time is, generally, not a ground state of the Hamiltonian $H$.

\section{Complex Wavefunctions}
\label{aq2c}

We now briefly suggest and elaborate on several extensions of our calculations thus far.
We will illustrate and suggest how our results may hold for general systems with 
complex wavefunctions. This will enable us
to go from a given quantum mechanical problem 
(including that of a fermionic system) to a corresponding classical
one. 

The similarity transformation of Eqs. (\ref{QMAC}, \ref{ssttate}, \ref{QMAC+}) captures a simple mathematical identity between
the generalized probability distribution of a classical system, obeying the Fokker-Planck dynamics with an operator $H_{FP}$ (Eq. (\ref{FPEQ})), and
the wavefunction obeying the Schr\"{o}dinger equation of the quantum dual Hamiltonian $H$ .
%The similarity transformation of Eqs. (\ref{QMAC}, \ref{ssttate}, \ref{QMAC+}) from a generalized probability distribution classical system obeying the Fokker-Planck dynamics with an operator $H_{FP}$ (Eq. (\ref{FPEQ})) to 
%a wavefunction obeying the Schr\"{o}dinger equation with the quantum Hamiltonian $H$ captures a simple mathematical identity.
Given this relation, it is possible to, formally, consider extensions in which the function evolving with the Fokker-Planck dynamics need not be 
a probability distribution. Most of our results concerning temporal correlations may hold under such a generalized interchange if we invoke
Eq. (\ref{psie}) to define (when given a quantum problem) a corresponding classical
system which need not be a bona fide physical Boltzmann distribution as in the scalar bosonic systems which we primarily focused on thus far and 
employ $|\Psi_{0}|^2$ as the initial (or final) time weight in the multiple time classical correlation function of 
Section \ref{TwoOperators}.
In Appendix \ref{sec:action}, we examined a formally imaginary counterpart to $V_{\sf N}$ and explicitly demonstrated how it leads to
standard results. Thus, given a quantum wavefunction, we may consider its logarithm to correspond to 
a classical potential $V_{\sf N}$. Wavefunctions of spinless Fermi systems cannot be purely positive
and for these complex (as well as divergent) potentials will formally arise. There may be subtleties however in our imaginary time ($t \to it$) analytic
continuations when $V_{\sf N}$ is not purely real (and the system effectively not purely dissipative) which are
more complex than those which we invoked thus far in our analysis of real $V_{\sf N}$ which led to response functions
of pure damped modes and their superpositions such as those which we encountered in Eq. (\ref{stretch}) [see also
Appendix \ref{dets}]. Physically, these are related to analogs of classical systems with instantons and tunneling events
(the behavior for the pure dissipative system) when these further exhibit non-damped oscillatory behavior.

%\bibliography{gcs}

\begin{thebibliography}{28}

\bibitem{subir_book}
S. Sachdev, {\it Quantum Phase Transitions}, (Cambridge University Press, 2004). 

\bibitem{physics_reports}
C. M. Varma, Z. Nussinov, and W. van Saarloos, Physics Rep. {\bf 361}, 267  (2002).

\bibitem{top} Y. S. Oh, \emph{et al}., Phys. Rev. Lett. {\bf  98}, 016401 (2007).

\bibitem{aji} V. Aji and C. M. Varma, Phys. Rev. B {\bf 82}, 174501 (2010).

\bibitem{nvb} Z. Nussinov, I. Vekhter, and A. V. Balatsky, Phys. Rev. B {\bf 79}, 165122 (2009).

\bibitem{She} J-H. She, J. Zaanen, A. R. Bishop, and A. V. Balatsky, Phys. Rev. B {\bf 82}, 165128 (2010).

\bibitem{montanari}
A. Montanari and G. Semerjian, J. Stat. Phys. {\bf 125}, 23 (2006). 

\bibitem{BiroliPointToSet}
G. Biroli, J.-P. Bouchaud, A. Cavagna, T. S. Grigera, and P. Verrocchio, 
Nature Phys. {\bf 4}, 771 (2008).

\bibitem{tanakaorder}
H. Tanaka, T. Kawasaki, H. Shintani, and K. Watanabe, Nature Mat. {\bf 9}, 324 (2010).

\bibitem{LawlerNematicOrder}
%M. J. Lawler, K. Fujita, Jhinhwan Lee, A. R. Schmidt, Y. Kohsaka, Chung Koo Kim, H. Eisaki,
%S. Uchida, J. C. Davis, J. P. Sethna, and Eun-Ah Kim, 
M. J. Lawler, \emph{et al}., Nature Lett. {\bf 466}, 347 (2010).

\bibitem{papa} C. Panagopoulos and V. Dobrosavljevic, Phys. Rev. B {\bf 72}, 014536 (2005).

\bibitem{physics_ZN}
Z. Nussinov, Physics {\bf 1}, 40 (2008); arXiv:1203.4648 (2012).

\bibitem{olmos}
B. Olmos, I. Lesanovsky, and J. P. Garrahan,
Phys. Rev. Lett. {\bf 109}, 020403 (2012).

\bibitem{htbib} L. Berthier and G. Biroli, Rev. Mod. Phys. {\bf 83}, 587 (2011).

\bibitem{EA}  S. F. Edwards and P. W. Anderson, J. Phys. F: Metal Phys. {\bf 5}, 965 (1975).

\bibitem{4p} C. Dasgupta  A. V. Indrani, S. Ramaswamy, and M. K. Phani, Europhys Lett.{\bf 15}, 307 (1991).

\bibitem{Parisi}
G. Parisi, {\it Statistical Field Theory}, (Addison Wesley, New York, 1988).

\bibitem{Zinn}
J. Zinn-Justin, {\it Quantum Field Theory and Critical Phenomena}, 
(Oxford University Press, 2002).

\bibitem{fenyes} I. Fenyes, Z. Phys. {\bf 132}, 81 (1952).

\bibitem{Nelson}
E. Nelson, Phys. Rev. {\bf 150}, 1079 (1966).

\bibitem{Guerra1}
F. Guerra and P. Ruggiero, Phys. Rev. Lett. {\bf 31}, 1022 (1973).

\bibitem{Guerra2}
F. Guerra and L. Morato, Phys. Rev. D {\bf 27}, 1774 (1983).

\bibitem{Biroli}
G. Biroli, C. Chamon, and F. Zamponi, Phys. Rev. B {\bf 78}, 224306 (2008).

\bibitem{castel}
C. Castelnovo, C. Chamon, and D. Sherrington, 
Phys. Rev. B {\bf 81}, 184303 (2010).

\bibitem{fei} M. V. FeigelÕman and M. A. Skvortsov, Nuc Phys B {\bf 506}, 665 (1997).

\bibitem{madelung} E. Madelung, Z. Phys. {\bf 40}, 322 (1926).

\bibitem{RK*} D. S. Rokhsar and S. A. Kivelson, Phys. Rev. Lett. {\bf 61}, 2376 (1988).

\bibitem{RK} Christopher L. Henley,  J. Phys. Condens. Matt.  {\bf 16}, S891 (2004); C. Castelnovo and C. Chamon, Ann. Phys. {\bf 318}, 316 (2005).

\bibitem{SQ} P. H. Damgaard and H. Huffel, Physics Reports {\bf 152}, 227 (1987); G. Parisi and Y.-S. Wu, Sci. Sinica {\bf 24}, 484 (1981).

\bibitem{holo} J. M. Maldacena, Adv. theor. Math. Phys. {\bf 2}, 231 (1998); 
{Int. J. Theor. Phys. 38, 1113 (1999); }
S. Gusber, I. R. Klebanov, and A. M. Polyakov, Phys. Lett. B {\bf 428}, 105 (1998); 
E. Witten, Adv. Theor. Math. Phys. {\bf 2}, 253 (1998);
O. Aharony, S. S. Guber, J. M. Maldacena, H. Ooguri, and Y. Oz, Phys. Rep. {\bf 323}, 183 (2000);
Z. Nussinov, G. Ortiz, and E. Cobanera, Annals of Physics {\bf 327}, 2491 (2012).


\bibitem{FPB} W. T. Coffey, Y. P.  Kalmykov, and J. T.  Waldron, {\it The Langevin Equation: With Applications to Stochastic Problems in Physics, Chemistry, and Electrical Engineering} (World Scientific, Singapore, 2004);
C. W. Gardiner, {\it Handbook of Stochastic Methods} (Springer-Verlag, Berlin, 2004).

\bibitem{Risken} H. Risken, {\it The Fokker-Planck Equation}, (Springer-Verlag, Berlin, 1996), Chap.\ 6.

\bibitem{SUSY} G. Junker, {\it Supersymmetric Methods in Quantum and Statistical Physics}, (Springer-Verlag, Berlin, 1996);
M.O. Hongler and W.M. Zheng, J. Stat. Phys. {\bf 29}, 317 (1982); M.O. Hongler and W.M. Zheng, J. Math. Phys. {\bf 24}, 336 (1983);
M. Bernstein and L.S. Brown, Phys. Rev. Lett. {\bf 52}, 1933 (1984); M. Hron and M. Razavy, J. Stat. Phys. {\bf 38}, 655 (1985);
H.R. Jauslin, J. Phys. A {\bf 21}, 2337 (1988); M.J. Englefield, J. Stat. Phys. {\bf 52}, 369 (1988).

\bibitem{forster} D. Forster, {\it Hydrodynamic Fluctuations, Broken Symmetry, and Correlation Functions} (New York: Benjamin, 1975)

\bibitem{repulsive} D.N. Perera and P. Harrowell, Phys. Rev. E {\bf 59}, 5721 (1999).

\bibitem{KA} W. Kob and H. C. Andersen, Phys. Rev. Lett. {\bf 73}, 1376 (1994); Phys. Rev. E {\bf 51}, 4626 (1995); {\bf 53}, 4143 (1995).

\bibitem{2DN3} Z. Rotman and E. Eisenberg, 
Phys. Rev. Lett. {\bf 105}, 225503 (2010).

\bibitem{3DN2}  H. Levit, Z. Rotman, and E. Eisenberg, Phys. Rev. E {\bf 85}, 011502 (2012).

\bibitem{BM} G. Biroli and M. Mezard, Phys. Rev. Lett. {\bf 88}, 025501(2002).

\bibitem{Phase1} 
K. S. Cole and R. H. Cole, J. of Chem. Phys. {\bf 9}, 341 (1941).
 
 \bibitem{Phase2}
66. D.W. Davidson and R. H. Cole, Journal of Chem. Phys. {\bf 18}, 1417 (1950); ibid. {\bf 19}, 1484
(1951).

\bibitem{Rault00} J. Rault, J. of Non-Crystalline Solids {\bf  271}, 177 (2000).

\bibitem{Johari}  G. P. Johari, and M. Goldstein, J. Chem. Phys. {\bf 53}, 2372 (1970).

\bibitem{alphabeta} M. Mierzwa, S. Pawlus, M. Paluch, E. Kaminska, and K. L. Ngai,
J. Chem. Phys. {\bf 128}, 044512 (2008); K. L. Ngai, Z. Wang, X. Q. Gao, H. B. Yu, and W. H. Wang, arXiv:1303.7424 (2013).


\bibitem{density_s} 
C. Dreyfus, A. Le Grand, J. Gapinski, W. Steffen, and A. Patkowski,
The European Physical Journal B {\bf 42}, 309 (2004)

\bibitem{MCT} D. R. Reichman and P. Charbonneau, J. Stat. Mech. P05013  (2005).

\bibitem{het.}
H. Sillescu, J. Non-Cryst. Solids {\bf 243}, 81 (1999);
M.D. Ediger, Ann. Rev. Phys. Chem.  {\bf 51}, 99 (2000);
R. Richert, J. Phys.: Condens. Mat.  {\bf 14}, R703 (2002);
W. Kob, C. Donati, S.J. Plimpton, P.H. Poole,
and S.C. Glotzer, Phys. Rev. Lett. {\bf 79}, 2827 (1997);
C. Donati, J.F. Douglas, W. Kob, S.J. Plimpton,
P.H. Poole, and S.C. Glotzer, Phys. Rev. Lett. {\bf 80}, 2338 (1998);
S.C. Glotzer, J. Non-Cryst. Solids {\bf 274}, 342 (2000);
Y. Gebremichael, T.B. Schroder, F.W. Starr, and
S.C. Glotzer, Phys. Rev. E {\bf 64}, 051503 (2001).


\bibitem{karmakar} S. Karmakar, C. Dasgupta, and S. Sastry, PNAS {\bf 106}, 3675 (2009).

\bibitem{wkob} W. Kob,  C. Donati, S. J. Plimpton, P. H. Poole, and S. C. Glotzer, 
Phys. Rev. Lett. {\bf 79}, 2827    (1997).

\bibitem{DK} R. P. A. Dullens and W. K. Kegel, Phys. Rev. E {\bf 71}, 011405 (2005).

\bibitem{lin} J.-X. Lin, C. Reichhardt, Z. Nussinov. L. P. Pryadko, and C. J. Olson Reichhardt, 
Phys. Rev. E {\bf 74}, 011403  (2006).

%\bibitem{kitaev} 
%A. Kitaev, Ann. Phys. {\bf 303}, 2 (2003).   
%  \bibitem{ref:montanariCL}
 % A. Montanari and G. Semerjian, Journal of Statistical Physics {\bf 125}, 23-54 (2006).
    
 \bibitem{ref:mosayebiCLSGT}    M. Mosayebi, E. D. Gado, P. Iig, and H. C. Ottinger,  Phys. Rev. Lett. {\bf 104}, 205704 (2010).
 
 \bibitem{ref:berthierCL} L. Berthier, G. Biroli, J.-P. Bouchaud, L. Cipelletti, D. El
Masri, D. L'Hote, F. Ladieu, and M. Pierno, Science {\bf 310}, 1797 (2005).

\bibitem{ref:karmakarsastry} S. Karmakar, C. Dasgupta, and S. Sastry,  Proc. Natl.
Acad. Sci. U.S.A. {\bf 106}, 3675 (2010).

\bibitem{ps} J.-P. Bouchaud and G. Biroli, J. Chem. Phys. {\bf 121}, 7347 (2004).

\bibitem{kl}
J. Kurchan and D. Levine,  e-print arXiv:0904.4850 (2009).

\bibitem{ref:aharonov} 
 E. Aharonov, E. Bouchbinder, H. G. E. Hentschel,
            V. Ilyin, N. Makedonska, I. Procaccia, and N. Schupper,
              Euro. Phys. Lett. {\bf 77}, 56002 (2007).

\bibitem{sheng} H. W. Sheng, W. K. Luo, F. M. Alamgir, J. M. Bai, and E. Ma, Nature {\bf 439}, 419 (2006) .

\bibitem{ref:finney} J. L. Finney, Proc. Roy. Soc. London, Ser. A {\bf 319}, 1539, 479 (1970).
           
           \bibitem{HA}  J. Dana Honeycutt and Hans C. Andersen, J. Phys. Chem. {\bf 91}, 4950 (1987).
           
           \bibitem{BO} P. J. Steinhardt, D. R. Nelson and M. Ronchetti,  Phys. Rev. B {\bf 28}, 784(1983).
           
           \bibitem{graph_method} P. Ronhovde, S. Chakrabarty, M. Sahu, K. F. Kelton, N. A. Mauro, K . K. Sahu, and Z. Nussinov,  European Physics Journal E  {\bf 34}, 105 (2011); P. Ronhovde, S. Chakrabarty, M. Sahu, K. K. Sahu, K. F. Kelton, N. Mauro, and Z. Nussinov,  Scientific Reports {\bf 2}, 329 (2012).

\bibitem{mizuno}  H. Mizuno and R. Yamamoto, Phys. Rev. E {\bf 84}, 011506 (2011).

\bibitem{J} A.J. Liu and S. R. Nagel, Nature (London) {\bf 396}, 21 (1998).

 \bibitem{arcm}  A.  J. Liu and S. R. Nagel, Annu. Rev. Condens. Matter Phys. {\bf 1}, 347 (2010).

\bibitem{J1} C. S. O'Hern, S. A. Langer, A. J. Liu, and S. R. Nagel, Phys. Rev. Lett. {\bf 88}, 075507 (2002).

\bibitem{J2} C. S. O'Hern, L. E. Silbert, A. J. Liu, and S. R. Nagel, Phys. Rev. E {\bf 68}, 011306 (2003).

\bibitem{J3} J. A. Drocco, M. B. Hastings, C. J. Olson Reichhardt, and  C. Reichhardt,
Phys. Rev. Lett. {\bf 95}, 088001 (2005).

\bibitem{silbert} L. E. Silbert, A. J. Liu, and S. R. Nagel, Phys. Rev. Lett. {\bf 95}, 098301 (2005).

\bibitem{J4} O. Dauchot, G. Marty, and G. Biroli, Phys. Rev. Lett. {\bf 95}, 265701 (2005).

\bibitem{J5} A. R. Abate and D. J. Durian, Phys. Rev. E {\bf 74}, 031308 (2006).

\bibitem{J6} A. R. Abate and D J. Durian, Phys. Rev. E {\bf 76}, 021306 (2007).

\bibitem{J7} A. S. Keys, A. R. Abate, S C. Glotzer, and D. J. Durian, Nature Phys. {\bf 3}, 260 (2007).

\bibitem{J8} F. Lechenault, O. Dauchot, G. Biroli, and J.-P. Bouchaud, Europhys. Lett. {\bf 83}, 46003 (2008).

\bibitem{hatano}  T. Hatano,  Phys. Rev. E {\bf 79}, 050301(R) (2009).

\bibitem{another4.6}  J. Lidmar and M. Wallin, Europhysics Letters {\bf 47}, 494 (1999)

\bibitem{zer} Z. Zeravcic, N. Xu, A. J. Liu, S. R. Nagel, and W. van Saarloos, Europhysics Letters {\bf 87}, 26001 (2009).

\bibitem{feld} B. U. Felderhof, Reports Math. Phys. {\bf 1}, 215 (1970); 
 B. U. Felderhof and M. Suzuki, Physica {\bf 56}, 43 (1971)
 
 \bibitem{reichhardt2004}
C. Reichhardt and C. J. Olson Reichhardt, Phys. Rev. Lett. {\bf 93},  176405 (2004)

\bibitem{ludovic} L. Berthier, H. Jacquin, and F. Zamponi, Phys. Rev. E {\bf 84}, 051103 (2011)

\bibitem{ludovic1} A. Ikeda, L. Berthier, and G. Biroli, J. Chem. Phys. {\bf 138}, 12A507 (2013)
   
%\bibitem{parisi} ``Spin Glass Theory
%and Beyond'', 
%M. Mezard, G. Parisi and
%M. A. Virasoro
%World Scientific, Singapore
%(Teaneck, N J.), 1987. 


\bibitem{mpa} M. P. A. Fisher, P. B. Weichman, G. Grinstein, and D. S. Fisher, Phys. Rev. B {\bf 40}, 546 (1989) 

\bibitem{zim} G.T. Zimanyi, P.A. Crowell, R.T. Scalettar, and G.G. Batrouni,
Phys. Rev. B {\bf 50}, 6515 (1994).

\bibitem{sw} J. Schmalian and P. G. Wolynes, Phys. Rev. Lett. {\bf 85},
836 (2000).

\bibitem{MM} T. Matsubara and H. Matsuda, Prog. Theor. Phys. {\bf 16}, 569
(1956).

\bibitem{BK} D. Belitz and T. R. Kirkpatrick, Phys. Rev. B {\bf 85}, 125126 (2012) 

\bibitem{Bulaevskii2011} L. N. Bulaevskii, M. J. Graf, C. D. Batista, and V. G. Kogan, Phys. Rev. B {\bf 83}, 144526 (2011).
\bibitem{Bulaevskii2012} L. N. Bulaevskii, M. J. Graf, and V. G. Kogan, Phys. Rev. B {\bf 85}, 014505 (2012).




\end{thebibliography}

\end{document}